\documentclass[12pt]{article} 
\usepackage{amsfonts}
\usepackage{amsmath}
\usepackage{amssymb}
\usepackage{amsthm}
\usepackage{array}
\usepackage[nosort]{cite}
\usepackage{caption}
\usepackage{color}
\usepackage{dsfont}
\usepackage{float}      
\usepackage{framed}
\usepackage{graphicx}
\usepackage{subcaption}
\usepackage{tabularx}
\usepackage{indentfirst}
\usepackage{mathrsfs}
\usepackage{multirow}
\usepackage{setspace}
\usepackage{braket}
\usepackage{pifont,makecell}
\usepackage[normalem]{ulem}
\usepackage{cancel}
\usepackage{subdepth}
\usepackage{titlesec}
\usepackage[dotinlabels]{titletoc}
\usepackage{wrapfig}
\usepackage{layout}
\usepackage{url}
\usepackage{slashed}
\usepackage{authblk}
\usepackage{tikz}
\usepackage{diagbox}
\usepackage[hyphens]{xurl}
\usepackage{lipsum}
\usepackage{etoolbox}
\usepackage{xcolor}
\usepackage{import}
\usepackage{mathtools}
\usepackage{example}
\usepackage{parskip}
\usepackage{chngcntr}
\usepackage[section]{placeins}
\usepackage{multicol}
\usepackage{makecell}
\usepackage[table]{xcolor}
\usepackage{nicematrix}
\NiceMatrixOptions{cell-space-limits=4pt}
\usepackage{booktabs}

\usepackage{comment}
\usepackage{hyperref}
\hypersetup{colorlinks=true}
\hypersetup{linkcolor=black}
\hypersetup{citecolor=blue}
\hypersetup{urlcolor=blue}

\numberwithin{equation}{section}

\usepackage[left=2.5cm,right=2.5cm,top=2.5cm,bottom=3cm]{geometry}
\def\tilde{\widetilde}

\def\hat{\widehat}

\def\bar{\overline}

\def\1{{\mathds 1}}

\newcommand{\Z}{{\mathbb Z}}

\def\CH{{\mathcal H}}

\def\CN{{\mathcal N}}
\def\CO{{\mathcal O}}

\newcommand{\eqa}[1]{\begin{align}#1\end{align}}
\newcommand{\eq}[1]{\begin{equation}#1\end{equation}}

\newcommand{\eqsp}[1]{\begin{equation}\begin{split}#1\end{split}\end{equation}}

\DeclareFontShape{OT1}{cmr}{mx}{n}%
    {<->cmr10}{}
\newcommand{\mytitlefont}{\fontseries{mx}\selectfont}
\DeclareMathAlphabet{\titlemath}{OT1}{cmr}{mx}{n}

\newcommand{\email}[1]{\vspace{5mm}\vbox{\center\footnotesize\tt#1}\vspace{5mm}}

\newcommand{\preprint}[1]{\begin{table}[t]  
             \begin{flushright}               
             {#1}                             
             \end{flushright}                 
             \end{table}}

\begin{document}

\preprint{CERN-TH-2026-270}

\begin{titlepage}

\begin{center}

~\\[10pt]

{\fontsize{30pt}{30pt} \mytitlefont  The Fate of 1/4-BPS States in the D1-D5 CFT \\ }

\vskip40pt

Suzanne Bintanja$^{\,a,\,b}$, Amey P. Gaikwad$^{\,c,\,b}$, Dikshant Rathore$^{\,b}$

\vskip20pt

{$^{a}$ \it CERN, Theory Division,
Geneva 23, CH-1211, Switzerland}\\[4pt]

{$^{b}$ \it Mani L.\,Bhaumik Institute for Theoretical Physics,\\[-2pt] Department of Physics and Astronomy,}\\[-2pt]
       {\it University of California, Los Angeles, CA 90095, USA}\\[4pt]

{$^{c}$ \it School of Physics and Astronomy, University of Minnesota,}\\[-2pt]
{\it Minneapolis, Minnesota 55455, USA}

\email{ suzanne.bintanja@cern.ch, agaikwad@umn.edu, drathore@physics.ucla.edu }

\vskip40pt

\end{center}

\begin{abstract}
    \noindent We consider the 1/4-BPS operators in the D1-D5 system on AdS$_3\times S^3\times T^4$. We evaluate the anomalous dimensions of low-lying sectors in conformal perturbation theory around Sym$^N(T^4)$ exactly in $N$. All anomalous dimensions are consistent with recombination of the short multiplets into long ones and take a particularly simple form. We show that at finite $N$ and at first nontrivial order in the deformation, for the sectors we consider, the BPS density of states agrees with that of the supergravity point. To the order in conformal perturbation theory we consider, this confirms that not all states that can get lifted do get lifted.  
\end{abstract}

\vfill

\end{titlepage}

\tableofcontents

\section{Introduction}

The canonical example of holography in the AdS$_3$/CFT$_2$ correspondence originates from the D1-D5 system \cite{Maldacena:1997re}. Type IIB supergravity on AdS$_3\times S^3\times M$, for $M\in\{T^4,\rm{K}3\}$, is dual to a point on the conformal manifold of the symmetric product orbifold of the nonlinear sigma model on $M$. The (grand-canonical ensemble of the) symmetric product orbifold of $T^4$, Sym$^N(T^4)$, is dual to a tensionless string theory with one unit of NS-NS flux \cite{Eberhardt:2018ouy,Eberhardt:2019ywk,Eberhardt:2020bgq,Gaberdiel:2020ycd}. The supergravity point is infinitely far removed from the orbifold on the conformal manifold. While we have a lot of mathematical control at the orbifold point, less is known about the CFT at more general points of the conformal manifold. Certain quantities are constant on the conformal manifold, providing a window to the microscopic description of gravity. The most notable example is the elliptic genus, a signed count of 1/4-BPS states, which is an index on the conformal manifold. In the case $M=T^4$, which preserves contracted large $\mathcal{N}=(4,4)$ supersymmetry, the elliptic genus vanishes due to fermionic zero modes.\footnote{There is however a modified elliptic genus that is nonzero \cite{Maldacena:1999bp}.} The 1/4-BPS Hilbert space is not constant; short representations $\chi_j$ of the $\mathcal{N}=(4,4)$ superconformal algebra can combine into long representations $\chi_{h,j}$ according to 
\eq{\label{eq:rec}
\chi_j+2\chi_{j+\frac{1}{2}}+\chi_{j+1}=\chi_{j,j}\,,
}
and acquire an anomalous dimension, removing them from the BPS subspace (see also Section~\ref{sec:shortrecomb}). The 1/4-BPS density of states can thus jump on the conformal manifold.

In recent years, there has been a lot of progress in understanding the microscopic properties of supersymmetric black holes. The BPS subspace (away from the orbifold point) can be separated into monotone operators, which are BPS for all values of $N$ and dual to a gas of supergravitons, and fortuitous operators, which are dual to black hole microstates and are not BPS above some critical value $N_\star$ \cite{Chang:2024zqi,Chang:2025rqy,Chang:2025wgo}. Furthermore, the chaotic properties of supersymmetric black holes can be probed by studying the statistics of the eigenspectrum of $\tilde{O}=P_{\text{BPS}}OP_{\text{BPS}}$, the projection of a simple operator $O$ onto the BPS subspace \cite{Lin:2022zxd,Lin:2022rzw,Chen:2024oqv}. 
Finally, BPS black holes satisfy a constraint on their charges. In the dual CFT, this implies that all black hole microstates satisfy the same constraint, which in this context takes the form $\tilde{j}=\frac{N}{2}$; the right-moving SU(2) R-charge is equal to $N/2$ \cite{Larsen:2021wnu,Larsen:2024fmp,Larsen:2025jqo}.

Each of the aforementioned phenomena is expected to emerge once we move on the conformal manifold towards the supergravity point due to the recombination and lifting of short multiplets. In fact, an enormous number of the BPS states have to lift away. At the orbifold point, the asymptotic light density of states of BPS operators exhibits Hagedorn growth in the large-$N$ limit \cite{Benjamin:2016pil}
\eq{
\rho^{\text{Sym}^N}_{\text{BPS}}(\Delta)\sim e^{2\pi\Delta}\,,
}
while in the supergravity theory it grows as
\eq{
\rho_{\text{BPS}}^{\text{sugra}}(\Delta)\sim e^{\Delta^{3/4}}\,.
}
In the black hole regime, it was recently argued using modular invariance and the asymptotic density of light states that an order one fraction of the $e^{S_0}$ BPS states, where $S_0$ is the entropy of the extremal black hole, has to lift from the BPS subspace by the time we reach the supergravity point \cite{Belin:2026vfd}. Furthermore, due to the nonrenormalization theorems of \cite{deBoer:2008ss,Baggio:2012rr}, which state that the OPE coefficients of `true' BPS operators (that are BPS everywhere) are covariantly constant on the conformal manifold, the only mechanism for BPS chaos to emerge is via lifting.\footnote{For the nonrenormalization theorems to apply, we must choose our simple operator $O$ to be BPS.} The idea is that the projection onto the BPS subspace can be understood as a projection onto a randomly oriented hypersurface \cite{Belin:2026ssc}. Understanding the recombination of short multiplets and the lifting mechanism on the conformal manifold is thus crucial to shed light on the properties of BPS black holes laid out above.

The central question we aim to answer is to what extent the supergravity spectrum is visible in perturbation theory around the orbifold point. Building on earlier results in conformal perturbation theory \cite{Burrington:2012yq,Gaberdiel:2015uca,Hampton:2018ygz,Guo:2019ady,Keller:2019suk,Guo:2020gxm,Apolo:2022fya,Guo:2022ifr,Hughes:2023apl,Hughes:2023fot,Gaberdiel:2026jor,Zhang:2026jnf}, and, in particular, generalizing the results of Benjamin, Keller and Zadeh (BKZ) \cite{Benjamin:2021zkn}, we study the lifting of low-lying 1/4-BPS operators away from the orbifold point. The pattern that emerged from these earlier works is that (at finite $N$) all states that can lift do indeed lift, already at first nontrivial order in perturbation theory. The supergravity spectrum is not consistent with this pattern. It contains short representations that can in principle recombine according to \eqref{eq:rec}. BKZ found evidence for non-maximal lifting in the $N\rightarrow\infty$ limit, but restricted the computation to just the untwisted sectors. To generalize the computation to finite $N$, mixing between the different twisted sectors has to be taken into account.

In this work, we evaluate the anomalous dimensions of 1/4-BPS operators to quadratic order in the deformation for the sectors with $h=1$, $j=0$, and $\tilde{h}=\tilde{j}\in\{0,\frac{1}{2},1\}$, exactly in $N$. The elements of the lifting matrix are computed as finite sums of OPE coefficients via the Gava-Narain method \cite{Gava:2002xb}. For $N\ge4$ all 1/4-BPS operators with $\tilde{j}\in\{0,\frac{1}{2}\}$ lift, while ten of the $\tilde{j}=1$ operators remain unlifted. These counts agree with the supergravity spectrum \cite{Benjamin:2017rnd}. We treat $N\in\{2,3\}$ separately. Including the twisted sectors leads to nontrivial mixing and results in anomalous dimensions of the simple form $\delta h=\delta\tilde{h}=\alpha\frac{\pi^2\lambda^2}{4(N-1)}$, with $\alpha\in\{\frac{1}{3},\frac{4}{9},\frac{1}{2},\frac{2}{3},1\}$. In particular, there are no anomalous dimensions scaling as $\mathcal{O}(\lambda^2/N^2)$, which would appear if we truncated the lifting matrix to just the untwisted sector. 

This paper is organized as follows. Section \ref{sec:setup} introduces our notation for Sym$^N(T^4)$ and contains a discussion on the $1/4$-BPS sector of the theory under consideration. In Section \ref{sec:deformation} we introduce the exactly marginal deformation used to deform the orbifold, and derive the formula we use to compute the anomalous dimensions, which takes the form of a finite sum over OPE coefficients of exchanged operators that contribute to the lifting matrix. Then, in Section \ref{sec:compute} we explain how to compute these OPE coefficients by going to a covering surface. Section \ref{sec:results} collects the results of the lifting computations, which we discuss together with future directions in Section \ref{sec:disc}. We collect various details and examples of the computations in the appendices.

\paragraph{Note added.} During the completion of this work, we learned that another group has independently reached a similar conclusion, namely that not all states that can get lifted do get lifted in the D1-D5 system. Their work is forthcoming \cite{haoyutoappear}.

\section{BPS States in the symmetric product orbifold of~\texorpdfstring{$\boldsymbol{T^4}$}{T4}}\label{sec:setup}

In this section, we introduce the symmetric product orbifold of~$T^4$ and its twisted sectors.
We then review the classification of BPS states at the orbifold point, in particular the~$1/4$-BPS states, whose fate away from the orbifold point is the subject of the later sections. The material in this section is standard, and we largely follow BKZ~\cite{Benjamin:2021zkn}. Our conventions for the superconformal algebra are collected in Appendix~\ref{22sca}. The states whose lifting we compute are listed in Table~\ref{tab:spectrum}.

\subsection{The seed theory and its symmetries}\label{sec:seed}
The moduli space of the D1-D5 CFT contains a distinguished locus where the theory is described by the symmetric product orbifold
\begin{equation}
    \text{Sym}^N(T^4) \equiv (T^4)^{\otimes N}/ S_N~, \qquad c = \widetilde{c} = 6 N~,
\end{equation}
where~$c,~\widetilde{c}$ are the left- and right-moving central charges, respectively. The seed theory is the~$\CN=(4,4)$ supersymmetric nonlinear sigma model on~$T^4$, with~$c=\widetilde{c}=6$. It is a free CFT with four real bosons and four real fermions in each chiral sector. We package the left-moving fields into two complex bosons and two complex fermions. These are denoted by
\begin{equation}
    \partial X^{(p)}~,\qquad \partial \bar{X}^{(p)}~,\qquad \Psi^{(p)}~,\qquad \bar{\Psi}^{(p)}~,\qquad p = 1,2~.
\end{equation}
The identical right-moving fields are denoted by~$\bar\partial X^{(p)}$ and~$\widetilde\Psi^{(p)}$. More generally, a tilde always denotes the right-moving counterpart of a left-moving quantity. We spell out the properties of the left-moving fields. The right-moving ones behave analogously. 

It is often convenient to use a uniform label~$\Psi^{(I)},~\partial X^{(I)}$ with~$I=1,\dots ,4$ where
\begin{equation}
    \Psi^{(3)}\equiv \bar{\Psi}^{(1)}~,\qquad \Psi^{(4)}\equiv \bar{\Psi}^{(2)}~,\qquad \partial X^{(3)} \equiv \partial \bar{X}^{(1)}~,\qquad \partial X^{(4)} \equiv \partial \bar{X}^{(2)}~.
\end{equation}
We write~$\bar I$ for the conjugate label, so that~$\bar 1 = 3$,~$\bar 2 = 4$, and vice versa. 
The fields have mode expansions in terms of the modes~$\partial X^{(p)}_m$ and~$\Psi^{(p)}_r$, which are given by
\begin{equation}\label{eq:s3-modes}
    \partial X^{(I)}(z) = \sum_{n\in\Z} \frac{\partial X^{(I)}_n}{z^{n+1}}~,\qquad \Psi^{(I)}(z) = \sum_{r\in\Z+\frac{1}{2}} \frac{\Psi^{(I)}_r}{z^{r+\frac{1}{2}}}~.
\end{equation}
Moreover, the modes satisfy
\begin{equation}\label{eq:s3-algebra}
    [\partial X_m^{(I)}, \partial X_{n}^{(J)}] = m\, \delta^{I\bar J}\delta_{m+n,0}~,\qquad \{\Psi_r^{(I)},\Psi_s^{(J)}\} = \delta^{I\bar J}\delta_{r+s,0}~,
\end{equation}
with hermitian conjugates given by
\begin{equation}\label{eq:s3-dagger}
    \big(\partial X^{(I)}_n\big)^\dagger = \partial X^{(\bar I)}_{-n}~,\qquad \big(\Psi^{(I)}_r\big)^\dagger = \Psi^{(\bar I)}_{-r}~.
\end{equation}
A mode with index~$-n<0$ raises the conformal weight by~$n$. Throughout the paper, states are in the NS sector, so that the fermion modes in~\eqref{eq:s3-modes} are half-integer. Ramond-sector fermions, with~$r\in\Z$, appear only as an auxiliary tool on the covering surfaces of Section~\ref{sec:compute}.

The seed theory enjoys a small~$\CN=(4,4)$ superconformal symmetry. On each side, the algebra is generated by the stress tensor~$T$, the four supercurrents~$G^{\alpha A}$, and~$SU(2)$ R-symmetry currents~$J^a$ at level~$k=\frac{c}{6}$. Here,~$\alpha=\pm$ labels the R-symmetry doublet, while~$A=1,2$ labels a doublet of the outer automorphism~$SU(2)$ of the algebra. We frequently use the combined label~$(\alpha A)$, and write the four supercurrents as~$G^{(+1)}$,~$G^{(-1)}$,~$G^{(+2)}$,~$G^{(-2)}$. Our conventions, together with the free-field realization on~$T^4$, are collected in Appendix~\ref{22sca}. We label states by their weights~$(h,\widetilde h)$, their R-symmetry spins~$(j,\widetilde j)$, and the eigenvalues~$(j^3,\widetilde j^{\,3})$ of~$J^3_0$ and~$\widetilde J^3_0$. The bosons are neutral under the R-symmetry. The fermions~$\Psi^{(1)},\Psi^{(2)}$ carry~$j^3=+\frac{1}{2}$, while~$\Psi^{(3)},~\Psi^{(4)}$ carry~$j^3=-\frac{1}{2}$. Two of the supercurrents play a distinguished role
\eqsp{\label{eq:s3-Gplus1}
    G^{(+1)} = \Psi^{(1)}\partial X^{(3)}+\Psi^{(2)}\partial X^{(4)}\,,\ \   G^{(-2)} = \Psi^{(3)}\partial X^{(1)}+\Psi^{(4)}\partial X^{(2)}\,,\ \  \big(G^{(+1)}_r\big)^\dagger = G^{(-2)}_{-r}~.
}
These are the supercurrents that enter the marginal deformation in Section~\ref{sec:deformation}. The remaining generators are given in Appendix \ref{22sca}.

Two further structures of the~$T^4$ theory will be important. First, the rotations of~$T^4$ give an~$SO(4) \simeq SU(2)_1 \times SU(2)_2$. In bispinor notation, the bosons transform as~$\partial X^{A\dot A}$ and the fermions as~$\Psi^{\alpha\dot A}$, where~$\dot A$ is a doublet index of~$SU(2)_1$ and~$A$ is a doublet index of~$SU(2)_2$. The dictionary to the labels~$(I)$ is given in~\eqref{eq:appA-bispinor}. The supercurrents~$G^{\alpha A}=\epsilon_{\dot A\dot B}\Psi^{\alpha\dot A}\partial X^{A\dot B}$ are~$SU(2)_1$ singlets, so~$SU(2)_1$ commutes with the superconformal algebra, while~$SU(2)_2$ acts on the bosons alone and realizes the outer automorphism.\footnote{We follow the naming of~\cite{Avery:2010er,Guo:2019ady}: $SU(2)_1$ rotates the fermions and the bosons, while~$SU(2)_2$ rotates only the bosons.} These groups are broken by the compactification of the torus~\cite{Avery:2010er,Guo:2019ady}, but they act on the oscillator states considered in this paper, which carry no momentum or winding, and preserve their correlation functions at the order in~$\lambda$ at which we work. The symmetry~$SU(2)_1$ organizes our results: because the deformation is~$SU(2)_1$ invariant all states of an~$SU(2)_1$ multiplet are lifted by the same amount.

The fermions additionally carry a flavor~$SU(2)$, generated by currents~$\hat{J}^a$ built from fermion bilinears
\begin{equation}\label{eq:s3-flavor}
    \hat J^{+} = \Psi^{(1)}\Psi^{(4)}~,\qquad \hat J^{-} = \Psi^{(2)}\Psi^{(3)}~,\qquad \hat J^{3} = \frac{1}{2}\Big(\Psi^{(1)}\Psi^{(3)}-\Psi^{(2)}\Psi^{(4)}\Big)~.
\end{equation}
It acts as~$SU(2)_1$ on the fermions alone and therefore does not commute with the supercurrents. It is a genuine symmetry at the orbifold point, but it is not part of the superconformal algebra. Moreover, the theory contains the four~$u(1)$ currents~$\partial X^{(I)}$ together with their superpartners~$\Psi^{(I)}$. Adjoining these to the small~$\CN=(4,4)$ algebra yields the contracted large~$\CN=(4,4)$ superconformal algebra~\cite{Sevrin:1988ew,Maldacena:1999bp}, which organizes the BPS spectrum of the~$T^4$ theory.\footnote{The name derives from the large~$\CN=(4,4)$ algebra, whose chiral half~$A_\gamma$ contains two~$SU(2)$ current algebras at levels~$k^{\pm}$ as well as a~$u(1)$~\cite{Sevrin:1988ew}. The contracted algebra arises as an In\"on\"u-Wigner contraction in which one of the levels is sent to infinity, degenerating the corresponding~$SU(2) \times U(1)$ currents into~$u(1)^4$~\cite{Sevrin:1988ew,Maldacena:1999bp}. The small~$\CN=(4,4)$ algebra is obtained by removing the free bosons and fermions altogether.}

\subsection{The orbifold and twisted sectors}
In the orbifold theory, each field additionally carries a copy label. For instance, we write~$\Psi^{(I),i}$ with~$i=1,\dots,N$. Operators of the orbifold theory are~$S_N$-invariant combinations of the operators of the~$N$ copies. Following BKZ, we construct them from a representative~$\check{\phi}$ whose first~$L$ tensor factors are non-trivial,
\begin{equation}\label{orbit}
    \check{\phi} = \phi^1 \otimes \dots \otimes \phi^L \otimes |0\rangle^{\otimes(N-L)}~,
\end{equation}
by summing over its~$S_N$-orbit,
\begin{equation}\label{eq:orbit_norm}
    \phi = \frac{1}{\sqrt{N!\, A_\phi}} \sum_{g\in S_N} g \check{\phi}~,\qquad A_\phi\equiv \sum_{g\in S_N} \langle g\check{\phi}|\check{\phi}\rangle ~,
\end{equation}
which is unit-normalized. Here~$\check\phi$ may also be a linear combination of terms of the form~\eqref{orbit}. We call states with a single non-trivial tensor factor (more generally, a single non-trivial cycle) single-trace, and states with several non-trivial tensor factors multi-trace.\footnote{The terminology is borrowed from large-$N$ gauge theories. As there, the connected correlation functions of unit-normalized single-trace operators are suppressed by inverse powers of $N$~\cite{Pakman:2009zz,Belin:2015hwa}.}

In the symmetric product orbifold, the generators of the corresponding algebras at~$c=6N$ are the single-trace operators~$\sum_{i=1}^N T^i$, etc. Note that for~$N>1$, the superconformal generators are independent fields, since bilinears of the single-trace~$\Psi^{(I)}$ and~$\partial X^{(I)}$ are multi-trace, and no longer coincide with~$T,~G^{\alpha A},~J^{a}$. This distinction matters for what follows: the R-symmetries~$J^a$ are part of the superconformal algebra and, hence, survive everywhere on the conformal manifold, whereas nothing protects the flavor currents~$\hat{J}^a$ away from the orbifold point. Indeed, they acquire an anomalous dimension \cite{Benjamin:2021zkn}.

Being an orbifold, the theory contains twisted sectors, labeled by the conjugacy classes of~$S_N$, i.e.~by cycle types. The corresponding twist fields~$\sigma_g$ are characterized by the monodromy
\begin{equation}
    \CO^{\,i} (e^{2\pi i}z)\, \sigma_g(0) = \CO^{\,g(i)}(z)\, \sigma_g(0)~,
\end{equation}
where~$\CO^{\,i}$ is any field of the~$i$-th copy, so that the~$n$ copies of the seed fields permuted by a single~$n$-cycle combine into fields with modes quantized in units of~$1/n$~\cite{Lunin:2000yv}. The representative~\eqref{orbit} of a twisted state is built on fixed cycles. For example,
\begin{equation}\label{eq:twisted-reps}
    \check\phi_{(2)}=\phi^{(12)}\otimes|0\rangle^{\otimes(N-2)}~,\qquad
    \check\phi_{(2,1)}=\phi^{(12)}\otimes\phi^{3}\otimes|0\rangle^{\otimes(N-3)}~,
\end{equation}
where~$\phi^{(12)}$ is a state in the twisted sector of the transposition~$(12)$ and~$\phi^3$ is an excited state of the third copy. We label a state by the twisted and excited cycles of its representative, so that~\eqref{eq:twisted-reps} are a twist-$2$ and a twist-$(2,1)$ state. Hence, twist-$2$ and twist-$3$ states are single-trace, while twist-$(2,1)$ and twist-$(2,1^2)$ states are multi-trace. The orbit sum~\eqref{eq:orbit_norm} gives the corresponding~$S_N$-invariant states, and its orbit factors are discussed in Section~\ref{sec:combinatorics}.

The ground states of the twist-$2$ sector have conformal dimensions equal to
\begin{equation}
    h=\widetilde h=4\cdot\frac{1}{16}+4\cdot\frac{1}{16}=\frac{1}{2}~,
\end{equation}
where the first term comes from the four twisted real bosons and the second from the four twisted real fermions. Since the twisted fermions have zero modes, the ground states are degenerate and are usually attributed to spin fields. There are sixteen of them, four for the left-movers times four for the right-movers. Four of them form a~$(\mathbf 2,\mathbf 2)$ of the left- and right-moving R-symmetry,
\begin{equation}
    \sigma^{\pm\pm}~,\qquad (j^3,\widetilde j^{\,3})=\big(\pm\tfrac{1}{2},\pm\tfrac{1}{2}\big)~.
\end{equation}
The remaining twelve are singlets of the left-moving or the right-moving R-symmetry, or of both, and are not~$\frac{1}{2}$-BPS.

The fields~$\sigma^{\pm\pm}$ have~$h=j$ and~$\widetilde h=\widetilde j$, and are~$\frac{1}{2}$-BPS. In particular,~$\sigma^{++}$ is a chiral and~$\sigma^{--}$ an anti-chiral primary, and we normalize them so that~$(\sigma^{--})^\dagger = \sigma^{++}$. Their superdescendants give the exactly marginal operators of Section~\ref{sec:deformation}. Since the lowest chiral primary of the twist-$n$ sector has~$h=j=\frac{n-1}{2}$ \cite{Lunin:2001pw}, twist-$2$ is the only twisted sector with single-trace exactly marginal operators.\footnote{For a seed theory with~$c\le 3$, there can be single-trace exactly marginal operators in higher twisted sectors as well~\cite{Apolo:2022fya}.} The twist fields are constructed using the bosonized spin fields we discuss in Section~\ref{sec:bosonization}.

\subsection{Short multiplets and recombination}\label{sec:shortrecomb}

 We now recall the representation theory of the contracted large~$\CN=(4,4)$ algebra in the NS sector for the right-movers. Throughout the paper, a primary state is defined with respect to this algebra, that is, a state annihilated by all positive modes of~$\widetilde T,\,\widetilde G^{\alpha A},\,\widetilde J^a$, and also by those of the single-trace free fields~$\sum_{i}\widetilde\Psi^{(I),i}$ and~$\sum_i\bar\partial X^{(I),i}$. The algebra implies the unitarity bound~$\widetilde h\geq \widetilde j$, see~\eqref{eq:appA-bound}. At~$\widetilde{c}=6N$, there is a finite family of short representations~$\widetilde{\chi}_{\widetilde{j}}$, built on primaries with
\begin{equation}
    \widetilde{h} = \widetilde{j} ~,\qquad \widetilde{j} = 0,~\frac{1}{2}~,\cdots,~\frac{N-1}{2}~,
\end{equation}
and long representations~$\widetilde{\chi}_{\widetilde{h},\widetilde{j}}$ with~$\widetilde{h} > \widetilde{j}$. At the unitarity threshold~$\widetilde{h}\rightarrow\widetilde{j}$, a long representation decomposes into short ones as\footnote{This holds for~$\widetilde j\leq\frac{N-3}{2}$. The long representations exist for~$\widetilde j\leq \frac{N-2}{2}$, and at the top value~$\widetilde j=\frac{N-2}{2}$ the last term in~\eqref{recomb} is absent.}
\begin{equation}\label{recomb}
    \widetilde{\chi}_{\widetilde{j},\widetilde{j}} = \widetilde{\chi}_{\widetilde{j}} + 2\widetilde{\chi}_{\widetilde{j}+\frac{1}{2}}+ \widetilde{\chi}_{\widetilde{j}+1}~.
\end{equation}
The weight of a short representation is fixed by its spin and is therefore protected. Conversely, short multiplets present at the orbifold point can pair up according to~\eqref{recomb} into long multiplets, whose weight is then free to change. This recombination is the mechanism by which BPS states are lifted, and it is precisely what our perturbative computation probes. It also leads to a sharp consistency check: the anomalous dimension of a long multiplet whose lowest component sits at~$\widetilde j$ must appear again at~$\widetilde j+\frac{1}{2}$ with twice the multiplicity, and at~$\widetilde j+1$ with the same multiplicity.

\subsection{The 1/4-BPS spectrum}
A primary can be short with respect to both the left- and the right-moving algebra, with respect to one of them, or with respect to neither. We call the first two cases
\begin{equation}
    \tfrac{1}{2}\text{-BPS} : ~ \chi_j~\widetilde{\chi}_{\widetilde{j}}~,\qquad \tfrac{1}{4}\text{-BPS} : ~ \chi_{h,j}~\widetilde{\chi}_{\widetilde{j}}~, \qquad \text{with}\quad  h>j~,
\end{equation}
where by convention the right-movers are short. The right-moving highest-weight component of a~$1/4$-BPS primary~$\phi$ thus satisfies
\begin{equation}\label{eq:s3-shortening}
    \widetilde{h}=\widetilde{j}\,=\widetilde j^{\,3} ~,\qquad \widetilde{G}_{-\frac{1}{2}}^{(+1)}\phi \,=\widetilde{G}_{-\frac{1}{2}}^{(+2)}\phi =0 ~,
\end{equation}
while on the left $h>j$. The two conditions in~\eqref{eq:s3-shortening} are equivalent, because the superconformal algebra implies
\begin{equation}\label{eq:s3-norm}
    \big\| \widetilde{G}_{-\frac{1}{2}}^{(+A)}\phi \big\|^2 = \big(\widetilde h-\widetilde j^{\,3}\big)\,\big\|\phi\big\|^2~,\qquad A=1,2~,
\end{equation}
for any primary~$\phi$, see~\eqref{eq:appA-bound}. The~$1/2$-BPS states are short on both sides and cannot lift. Their non-lifting serves as a valuable consistency check of our computation in Section~\ref{sec:results}.

In this paper, we study the lifting of the~$1/4$-BPS primaries with~$(h,j)=(1,0)$ and~$\widetilde{j}\in\{0,\frac{1}{2},1\}$. Their degeneracies at the orbifold point are listed in Table~\ref{tab:spectrum}, which refines Table 1 of BKZ~\cite{Benjamin:2021zkn}. They are split into untwisted and twisted sectors, and listed together with the number of states with the same quantum numbers at the supergravity point~\cite{Benjamin:2016pil,Benjamin:2017rnd}. The fact that the supergravity numbers are much smaller than those at the orbifold point, yet larger than the minimal spectrum consistent with~\eqref{recomb}, is what makes the fate of these states a non-trivial dynamical question.

\begin{table}[ht]
\centering
\renewcommand{\arraystretch}{1.3}
\begin{tabular}{c||c|c|c|c|c}
$\widetilde{j}$ & sector & primaries & $SU(2)_1$ & minimal & supergravity\\
\hline
\hline
$0$ & untwisted & $3$ & $\mathbf 3$ & $0$ & $0$\\
\hline
\multirow{3}{*}{$\frac{1}{2}$} & untwisted & $6$ & $\mathbf 4\oplus\mathbf 2$ & \multirow{3}{*}{$0$} & \multirow{3}{*}{$0$}\\
 & twist-2 & $6$ & $\mathbf 3\oplus\mathbf 3$ & & \\
 & twist-$(2,1)$ & $2$ & $\mathbf 2$ & & \\
\hline
\multirow{4}{*}{$1$} & untwisted & $9$ & $\mathbf 3\oplus\mathbf 3\oplus\mathbf 3$ & \multirow{4}{*}{$3$} & \multirow{4}{*}{$10$}\\
 & twist-3 & $7$ & $\mathbf 3\oplus\mathbf 2\oplus\mathbf 2$ & & \\
 & twist-$(2,1)$ & $16$ & $2\cdot\mathbf 4\oplus\mathbf 3\oplus 2\cdot\mathbf 2\oplus\mathbf 1$ & & \\
 & twist-$(2,1^2)$ & $4$ & $\mathbf 3\oplus\mathbf 1$ & & \\
\end{tabular}
\caption{The~$1/4$-BPS primaries with~$(h,j)=(1,0)$ and~$\widetilde h=\widetilde j\leq 1$ at the orbifold point, for~$N\geq 4$. There are~$3$,~$14$ and~$36$ of them at~$\widetilde j=0,\frac{1}{2},1$ respectively. We count primaries of the contracted large~$\CN=(4,4)$ algebra, and only the highest-weight component~$\widetilde j^{\,3}=\widetilde j$ of each right-moving multiplet. The fourth column lists the multiplets of~$SU(2)_1$ in each sector. The column `minimal' is taken from Table 1 of BKZ~\cite{Benjamin:2021zkn}. It belongs to a spectrum with the smallest total number of~$1/4$-BPS primaries that is compatible with~\eqref{recomb}, assuming that all~$\widetilde j=0$ primaries are lifted. This spectrum is not unique, and it also involves primaries with~$\widetilde j>1$ not shown here. The last column denotes the number of such primaries at the supergravity point.}
\label{tab:spectrum}
\end{table}

The simplest primaries are those at~$\widetilde j=0$. They are purely left-moving and are built using the flavor currents~\eqref{eq:s3-flavor}. The single-trace current~$\sum_i\hat J^{a,i}$ is not, by itself, a primary, since it overlaps with the bilinear of the single-trace fermions, which is a descendant of the vacuum. Subtracting this bilinear gives the primary
\begin{equation}\label{eq:s3-flavorprimary}
    \phi^{+} = \frac{1}{\sqrt{N-1}}\bigg(\sum_{i=1}^N \Psi^{(1),i}_{-\frac{1}{2}}\Psi^{(4),i}_{-\frac{1}{2}} - \frac{1}{N}\sum_{i,j=1}^N \Psi^{(1),i}_{-\frac{1}{2}}\Psi^{(4),j}_{-\frac{1}{2}}\bigg)|0\rangle~,
\end{equation}
and similar constructions yield the primaries~$\phi^-$ and~$\phi^3$. The state~$\phi^+$ is the orbit~\eqref{eq:orbit_norm} of the representative~$\check\phi^+$ of Appendix~\ref{sec:example}. The three primaries~$\phi^a$ form a triplet of~$SU(2)_1$. The representatives of the primaries at~$\widetilde j=\frac12$ and~$\widetilde j=1$ are given in Appendix~\ref{app:sectors}.

Two features of the table deserve emphasis.
\begin{itemize}
     \item Most of the primaries lie in twisted sectors. Primaries with the same quantum numbers in different sectors are exactly degenerate at the orbifold point, and the deformation mixes them. Hence the lifting matrix of Section~\ref{sec:deformation} is defined on the full degenerate subspace, untwisted and twisted states alike. The~$S_N$ selection rules of Section~\ref{sec:combinatorics} allow mixing only between sectors whose permutations have the same parity. At~$\widetilde j=1$, the untwisted primaries can thus mix with the twist-$3$ ones, and the twist-$(2,1)$ primaries with the twist-$(2,1^2)$ ones. In~\cite{Benjamin:2021zkn}, the lifting matrix was restricted to its untwisted block, and including the twisted sectors is the main extension of this paper. Since the anomalous dimensions are the eigenvalues of the full matrix, the untwisted truncation is not sufficient at finite~$N$, nor in the large-$N$ expansion; see Section~\ref{sec:results-one}.

    \item The table lists the stable degeneracies, valid for~$N\geq 4$. For smaller $N$, the spectrum truncates, since the right-moving short representations only exist for~$\widetilde{j}\leq \frac{N-1}{2}$. Thus, for~$N=2$ only the families~$\widetilde{j}=0,~\frac{1}{2}$ occur, and for~$N=3$, only~$\widetilde{j}\leq 1$. Moreover, the degeneracies at small~$N$ differ from their stable values. The spectra and lifts at~$N=2$ and~$N=3$ are given in Section~\ref{sec:smallN}.

 \end{itemize}

\section{Deformation away from the orbifold point}\label{sec:deformation}

In this section, we review salient details concerning the conformal manifold of symmetric product orbifold theories and how to study it using conformal perturbation theory. Furthermore, we derive the formula we will use to compute anomalous dimensions.

\subsection{Conformal manifold}

The structure of symmetric product orbifold CFTs allows for interesting conformal manifolds. If the seed theory is sufficiently supersymmetric, the orbifold theory can contain exactly marginal operators that can be used to deform the theory. Such exactly marginal operators are given by the (Hermitian combination of) superdescendants of 1/2-BPS operators with $h=\tilde{h}=\frac{1}{2}$
\eq{\label{eq:mod}
O_\text{m}\equiv\frac{1}{\sqrt{2}}\left(\widetilde{G}_{-\frac{1}{2}}^{(+1)}G_{-\frac{1}{2}}^{(+1)}O_{\frac{1}{2}\text{-BPS}}+G_{-\frac{1}{2}}^{(-2)}\widetilde{G}_{-\frac{1}{2}}^{(-2)}O_{\frac{1}{2}\text{-BPS}}^{\dagger}\right)\,,
}
with $O_{\frac{1}{2}\text{-BPS}}$ an anti-chiral primary of $h=\tilde{h}=j=\tilde{j}=\frac{1}{2}$.
A marginal deformation is then given by adding the following term to the action
\eq{\label{eq:def}
S\mapsto S+\lambda\int d^2z \ O_\text{m}(z,\bar{z})\,.
}

There are two necessary conditions for the deformation to possibly lead to a supergravity point on the conformal manifold. First, in order to break the symmetry structure of the $S_N$ orbifold, $O_{\frac{1}{2}\text{-BPS}}$ has to be in the twisted sector. Second, in order for most operators to be able to obtain large anomalous dimensions, $O_{\frac{1}{2}\text{-BPS}}$ has to be single-trace. This can be seen by analyzing the large-$N$ scaling of correlation functions computed in conformal perturbation theory around the orbifold point \cite{Belin:2020nmp,Apolo:2022pbq,Apolo:2024bmu}. In the $T^4$ theory under consideration, this means that the deformation is the superpartner of the twist-2 ground state, i.e., $O_{\frac{1}{2}\text{-BPS}}=\sigma^{--}$.\footnote{This is a consequence of the fact that the seed central charge satisfies $c=6$. Single-trace twisted-sector deformations can be present when the seed central charge satisfies $1\le c\le 6$ \cite{Belin:2020nmp}.}

In this work, we are interested in the anomalous dimensions of $1/4$-BPS operators under the twisted sector single-trace deformation. These can be found by computing two-point functions in the deformed theory \eqref{eq:def} via conformal perturbation theory. More precisely, the normalized two-point function of ($1/4$-BPS) operators $\phi_{\ell,k}$ can be expanded as\footnote{When possible, we suppress the anti-holomorphic coordinates for simplicity.}
\eqa{
\langle\phi^\dag_\ell(x)\phi_k(y)\rangle_\lambda\equiv&\frac{\langle\phi^\dag_\ell(x)\phi_k(y)e^{-\lambda\int d^2z\, O_\text{m}(z)}\rangle}{\langle e^{-\lambda\int d^2z \,O_\text{m}(z)}\rangle}\notag\\
=&\langle\phi^\dag_\ell(x)\phi_k(y)\rangle-\lambda\int d^2z \langle \phi^\dag_\ell(x) O_\text{m}(z) \phi_k(y)\rangle_{\text{connected}}\\
&\hspace{40pt} +\frac{\lambda^2}{2}\int d^2z\, d^2z'\langle\phi^\dag_\ell(x) O_\text{m}(z) O_\text{m}(z')\phi_k(y)\rangle_{\text{connected}}+\mathcal{O}\left(\lambda^3\right)\,.\notag
}
Since~$\phi_{k,\ell}$ saturate the right-moving unitarity bound, the linear term has to vanish,\footnote{In fact, all odd powers of $\lambda$ vanish due to selection rules on twisted sector operators.} otherwise $\lambda$ can be tuned such that the right-moving dimension of $\phi_{k,\ell}$ violates the unitarity bound. The first correction to the conformal weights thus comes at second order in the deformation parameter~$\lambda$.

Because the $1/4$-BPS operators are generically degenerate, we need to take into account operator mixing. To this end, we define the lifting matrix
\eq{\label{eq:gammadef}
\gamma^{k\ell} \equiv -\frac{\pi}{2}\lambda^2 \int d^2x \ \mathcal{G}^{k\ell}(x,\bar{x})\,,
}
with 
\eq{
\mathcal{G}^{k\ell}(x,\bar{x})\equiv \langle \phi^\dagger_\ell(\infty,\infty) O_\text{m}(1,1)O_\text{m}(x,\bar{x})\phi_k(0,0)\rangle\,.
}
We will denote the anomalous dimensions as the eigenvalues of the lifting matrix, since each eigenvalue is the shift $\delta h=\delta\tilde h$ of the conformal weights, and the eigenvectors encode operator mixing under the deformation.\footnote{Strictly speaking the anomalous dimension $\delta\Delta=\delta h+\delta\tilde{h}=2\delta{h}$. We will slightly abuse the nomenclature and say that $\delta h$ is the anomalous dimension.} Only the cross terms in the product of the two insertions of \eqref{eq:mod} can contribute. In other words,
\eq{
\mathcal{G}^{k\ell}(x,\bar{x})=\frac{1}{2}\left(\mathcal{G}^{k\ell}_1(x,\bar{x})+\mathcal{G}^{k\ell}_2(x,\bar{x})\right)\,, 
}
where, with shorthand $O\equiv \widetilde{G}_{-\frac{1}{2}}^{(+1)}G_{-\frac{1}{2}}^{(+1)}\sigma^{--}$,
\eqsp{\label{eq:G12}
\mathcal{G}^{k\ell}_1(x,\bar{x})&=\langle \phi^\dagger_\ell(\infty,\infty) O^\dag(1,1)O(x,\bar{x})\phi_k(0,0)\rangle\,,\\
\mathcal{G}^{k\ell}_2(x,\bar{x})&=\langle \phi^\dagger_\ell(\infty,\infty) O(1,1)O^\dagger(x,\bar{x})\phi_k(0,0)\rangle\,.
}
Using superconformal Ward identities, we can reduce the integral \eqref{eq:gammadef} and the expressions in \eqref{eq:G12} to finite sums over OPE coefficients. This can be understood as selecting the exchanges that contribute to the integral in the OPE limit of the four-point functions above. We follow the framework introduced by BKZ \cite{Benjamin:2021zkn}, which is a slight generalization of the well-known Gava-Narain formula \cite{Gava:2002xb}.

\subsection{Decomposition in OPE coefficients}

The~$1/4$-BPS states saturate the BPS bound on the right, so that~$\tilde{G}_{-\frac{1}{2}}^{(+1)}\phi_{k,\ell}=0$ by~\eqref{eq:s3-shortening}. We also write
\eq{
O' = G^{(+1)}_{-\frac{1}{2}} \sigma^{--}\,,\qquad{\text{such that}}\qquad O = \tilde{G}^{(+1)}_{-\frac{1}{2}} O'\,,\qquad O^\dagger = \tilde{G}^{(-2)}_{-\frac{1}{2}} O'^\dagger\,.
}
Finally, we define
\eq{
   \mathcal{C}^{k\ell} (z_1,z_2,z_3,z_4) = \langle \phi_\ell^{ \dag}(z_1) O(z_2) O^\dag(z_3) \phi_{k}(z_4) \rangle\,.
}
From the superconformal algebra, it follows that acting with a supercurrent on the superdescendant $O$ gives rise to a total derivative (see Appendix C of
\cite{Keller:2019suk}). More precisely,
\eq{    
\oint_{z_2}\frac{d\bar z}{2\pi i}\,
 \bar\xi(\bar z)\,
 \widetilde G^{(-2)}(\bar z)\,
 O(z_2)
 =
 \partial_{\bar z_2}
 \left[
 \bar\xi(\bar z_2)\,O'(z_2)
 \right]\,.
 }
We choose
\eq{
\bar{\xi}(\bar{z}) = \frac{\bar z - \bar z_4}{\bar z_3 - \bar z_4}\,,
}
so that
\eq{
\mathcal{C}^{k\ell} (z_1,z_2,z_3,z_4) = - \partial_{\bar z_2} \left[ \frac{\bar z_2 - \bar z_4}{\bar z_3 - \bar z_4} \left\langle \phi_\ell^{ \dag}(z_1)\, O'^\dag(z_3)\, O'(z_2)\, \phi_{k}(z_4) \right\rangle  \right ]\,.
}
If we then assign
\eq{\label{eq:zpos}
z_1 = \infty, \quad z_2 = x, \quad z_3 = 1, \quad z_4 = 0\,,
}
we can write $\mathcal{G}_1^{k\ell}$ as a total derivative,
\eq{\label{eq:g1d}
    \mathcal G^{k\ell}_1(x,\bar x)
 =
 -\partial_{\bar x}
 I^{k\ell}_1(x,\bar x)\,,
}
with
\eq{
    I^{k\ell}_1(x,\bar x)
 =
 \bar x\,
 \left\langle
 \phi_\ell^{\dagger}(\infty)\,
  O'^\dagger(1)\,
  O'(x,\bar x)\,
 \phi_k(0)
 \right\rangle \,.
}
We can perform a similar argument to write $\mathcal{G}_2^{k\ell}$ as a total derivative. We now use
\eq{
    \oint_{z_3}\frac{d\bar z}{2\pi i}\,
 \bar\xi(\bar z)\,
 \widetilde G^{(+1)}(\bar z)\,
 O^\dagger(z_3)
 =
 \partial_{\bar z_3}
 \left[
 \bar\xi(\bar z_3)\,O'^\dagger(z_3)
 \right]\,,
}
with
\eq{
    \bar{\xi}(\bar z)
    =
    \frac{\bar z-\bar z_1}
         {\bar z_2-\bar z_1}\,,
}
to remove~$\tilde{G}_{-\frac{1}{2}}^{(-2)}$ from the insertion~$O^\dagger$ at~$z_3$. This leads to
\eq{
    \mathcal{C}^{k\ell} (z_1,z_2,z_3,z_4) = - \partial_{\bar z_3} \left[ \frac{\bar z_3 - \bar z_1}{\bar z_2 - \bar z_1} \left\langle \phi_\ell^{ \dag}(z_1)\, O'(z_2)\, O'^\dag(z_3)\, \phi_{k}(z_4) \right\rangle  \right ]\,.
}
Assigning 
\eq{
z_1 = \infty, \quad z_2 = 1, \quad z_3 = x, \quad z_4 = 0\,,
}
we find
\eq{\label{eq:g2d}
    \mathcal G^{k\ell}_2(x,\bar x)
 =
 -\partial_{\bar x}
 I^{k\ell}_2(x,\bar x)\,,
}
with
\eq{
     I^{k\ell}_2(x,\bar x)
 =
 \left\langle
 \phi_\ell^{\dagger}(\infty)\,
 O'(1)\,
 O'^\dagger(x,\bar x)\,
 \phi_k(0)
 \right\rangle .
}
Using \eqref{eq:g1d}, \eqref{eq:g2d} and Stokes' theorem, we can perform the integral over $\bar{x}$ in \eqref{eq:gammadef}. This reduces the integral to a contour integral over $x$, which can only receive contributions from the OPE limits $x\rightarrow 0$ and $x\rightarrow\infty$,\footnote{The OPE channel $x\rightarrow1$ does not contribute to the anomalous dimensions due to the fact that it comes from taking the OPE of the two moduli, see e.g. \cite{Keller:2019suk} for details.} 
\eqa{\label{eq:stokes}
\gamma^{k\ell}=-\frac{i \pi}{4}\lambda^2\int dxd\bar{x}\ \mathcal{G}^{k\ell}(x,\bar{x})&=\frac{i \pi}{8}\lambda^2\oint_{0,\infty} dx\ \left(I_1^{k\ell}(x,\bar{x})+I_2^{k\ell}(x,\bar{x})\right)\,.
}
The contour integral now only picks up the terms in the respective OPE expansions around $x=0$ and $x=\infty$ that come with the powers $x^{-1}\bar{x}^0$. This provides a restriction on the scaling dimension of the exchanged operators. For future convenience, before considering the OPE expansions, we will permute the operators in \eqref{eq:stokes} so that the $O'$ insertions are located at $z=0,\infty$. To do so, in the four-point functions we change coordinates according to
\eq{
w = \frac{x-z}{1-z}\,, \qquad (\infty,x,1,0) \longmapsto(1,0,\infty,x)\,.
}
We obtain
\eq{\label{eq:gammahat}
\gamma^{k\ell}=\frac{i\pi}{8}\lambda^2\oint_{0,\infty} dx\ \frac{(1-x)^{2h}(1-\bar x)^{2\tilde j}}
         {(x-1)^2(\bar x-1)}\left(\hat{I}_1^{k\ell}(x,\bar{x})+\hat{I}_2^{k\ell}(x,\bar{x})\right)\,,
}
with
\eqsp{
 \hat{I}^{k\ell}_1(x,\bar x)
 &=\bar{x}
 \left\langle
 O'^\dagger(\infty)\, 
 \phi_\ell^{\dagger}(1)\,\phi_k(x,\bar{x}) \,
O'(0)
 \right\rangle\,,\\
\hat{I}^{k\ell}_2(x,\bar x)
 &=
 \left\langle
 O'(\infty) \,
\phi_\ell^{\dagger}(1)\,
 \phi_k(x,\bar{x})\,
 O'^\dagger(0)
 \right\rangle\,.
}
Now we take the OPE limit $x\rightarrow 0$. The relevant OPEs are
\eq{
\phi_k(x,\bar{x})\,\mathcal{O}(0)\sim\sum_\chi C_{\chi^\dagger\phi_k\mathcal{O}}\,x^{h_\chi-h-1}\bar{x}^{\tilde{h}_\chi-\tilde{j}-\frac{1}{2}}\chi(0)\,,\qquad \mathcal{O}\in\{O',O'^\dagger\}\,.
}
Hence, by inserting a complete set of states $\chi$, we have
\eqsp{
\hat{I}_1^{k\ell}(x,\bar{x})\vert^0&=\sum_\chi C_{O'^\dagger\phi_\ell^\dagger \chi}C_{\chi^\dagger \phi_k O'}\,x^{h_\chi-h-1}\bar{x}^{\tilde{h}_\chi-\tilde{j}+\frac{1}{2}}\,,\\
\hat{I}_2^{k\ell}(x,\bar{x})\vert^0&=\sum_\chi C_{O'\phi_\ell^\dagger \chi}C_{\chi^\dagger \phi_k O'^\dagger}\,x^{h_\chi-h-1}\bar{x}^{\tilde{h}_\chi-\tilde{j}-\frac{1}{2}}\,.
}
The limit $x\rightarrow\infty$ is the OPE of $\phi_k$ with the insertion at infinity. Since the map $x\mapsto1/x$ exchanges $\hat I_1$ and $\hat I_2$ up to conformal factors, this results in\footnote{Here, with a small abuse of notation, we include the Jacobian in \eqref{eq:gammahat}.}
\eqsp{
\hat{I}_1^{k\ell}(x,\bar{x})\vert^\infty&=-\sum_\chi C_{O'\phi_\ell^\dagger \chi}C_{\chi^\dagger \phi_k O'^\dagger}\,x^{h-h_\chi-1}\bar{x}^{\tilde{j}-\tilde{h}_\chi+\frac{1}{2}}\,,\\
\hat{I}_2^{k\ell}(x,\bar{x})\vert^\infty&=-\sum_\chi C_{O'^\dagger\phi_\ell^\dagger \chi}C_{\chi^\dagger \phi_k O'}\,x^{h-h_\chi-1}\bar{x}^{\tilde{j}-\tilde{h}_\chi-\frac{1}{2}}\,.
}
Combining these expressions and picking out the powers of $x$ that survive the contour integral, we end up with
\eqsp{\label{eq:gamma_final}
\gamma^{k\ell}&=\frac{\pi^2\lambda^2}{2}\sum_{\substack{\chi:h_\chi=h\\ \tilde{h}_\chi=\tilde{j}+\frac{1}{2}}}C_{O'\phi_\ell^\dagger \chi}C_{\chi^\dagger \phi_k O'^\dagger}+\frac{\pi^2\lambda^2}{2}\sum_{\substack{\chi:h_\chi=h\\ \tilde{h}_\chi=\tilde{j}-\frac{1}{2}}}C_{O'^\dagger\phi_\ell^\dagger \chi}C_{\chi^\dagger \phi_k O'}\\
&=\frac{\pi^2\lambda^2}{2}\sum_{\substack{\chi:h_\chi=h\\ \tilde{h}_\chi=\tilde{j}+\frac{1}{2}}}C_{O'\phi_\ell^\dagger \chi}\left(C_{O' \phi_k^\dagger \chi}\right)^*+\frac{\pi^2\lambda^2}{2}\sum_{\substack{\chi:h_\chi=h\\ \tilde{h}_\chi=\tilde{j}-\frac{1}{2}}}\left(C_{\chi^\dagger\phi_\ell O'}\right)^*C_{\chi^\dagger \phi_k O'}\,.
}
Hence, the elements of the lifting matrix can be computed as a finite sum over OPE coefficients. The sum runs over all intermediate states~$\chi$ with~$h_\chi=h$ and~$\widetilde h_\chi=\widetilde j\pm\frac{1}{2}$, and both primaries and descendants can contribute. The next section explains the computation of these OPE coefficients.

\section{Computing twisted 3-point functions}\label{sec:compute}

Section \ref{sec:deformation} reduced the lifting matrix to a finite sum over three-point coefficients. In this section, we explain how these coefficients are computed. In general, both the $1/4$-BPS primaries and the intermediate states $\chi$ can live in different twisted sectors. The steps below are implemented in \texttt{Mathematica}, exactly in $N$. 
We compute the OPE coefficients given below,
\begin{align}
    C_{\chi^\dagger \phi_k O'} \equiv \big\langle \chi^\dagger(\infty)\, \phi_k(1)\, O'(0) \big\rangle\,,
    \qquad
    C_{\chi^\dagger \phi_k O'^\dagger} \equiv \big\langle \chi^\dagger(\infty)\, \phi_k(1)\, O'^\dagger(0) \big\rangle\,.
\end{align}
The intermediate states are summed over a basis
\begin{align}\label{eq:Hpm}
    \CH_\pm = \Big\{ \chi \,:\, h_\chi = h\,,\ \widetilde h_\chi = \widetilde j \pm \tfrac12 \Big\}\,,
\end{align}
restricted to the charges allowed by R-charge conservation. The $\chi$ states are not, in general, orthonormal, so we use their Gram matrices $(G_\pm)_{\chi\chi'} = \langle \chi | \chi' \rangle$, which gives
\begin{align}\label{eq:master_GN}
\begin{split}
    \gamma^{k\ell}
    =&
    \frac{\pi^2\lambda^2}{2} \sum_{\chi,\chi' \in \CH_+}
    \big( C_{\chi^\dagger \phi_\ell O'^\dagger} \big)^*
    \big( G_+^{-1} \big)_{\chi\chi'}\,
    C_{\chi'^\dagger \phi_k O'^\dagger}\\
    &\hspace{80pt}
    \;+\;
    \frac{\pi^2\lambda^2}{2} \sum_{\chi,\chi' \in \CH_-}
    \big( C_{\chi^\dagger \phi_\ell O'} \big)^*
    \big( G_-^{-1} \big)_{\chi\chi'}\,
    C_{\chi'^\dagger \phi_k O'}\,.
    \end{split}
\end{align}
We will call the two terms the $O'^\dagger$ channel and the $O'$ channel, respectively. For an orthonormal basis, \eqref{eq:master_GN} reduces to the sum of squares \eqref{eq:gamma_final}.

Each lifting matrix is computed via the following steps.
\begin{enumerate}
    \item We construct the $1/4$-BPS primaries with the required weights and charges as fixed-copy representatives. In a twisted sector, the representative is given via operator modes acting on the cover ground state of a fixed cycle. We then normalize the full $S_N$ orbit as in~\eqref{eq:orbit_norm}.

    \item We construct a complete basis of $\CH_\pm$, keeping only the twist sectors allowed by the selection rules discussed in Section~\ref{sec:combinatorics}, and compute its Gram matrix.

    \item For each assignment of copies, we lift the correlator to its covering space. There, we evaluate it in the bosonized free theory, including the local-coordinate factors, descendant corrections, if any, and the Liouville factor (as we discuss in Sections~\ref{sec:cover} and~\ref{sec:bosonization}).

    \item We assemble \eqref{eq:master_GN} from the coefficients and the Gram matrices. As an independent check, we also computed the coefficients $C_{O'\phi_\ell^\dagger\chi}$ and $C_{O'^\dagger\phi_\ell^\dagger\chi}$ of the first line of~\eqref{eq:gamma_final} directly, rather than as complex conjugates. The matrix $\gamma$ assembled from the two independently computed sets is Hermitian, providing a non-trivial consistency check.
\end{enumerate}

\subsection{Cover maps and local coordinates}\label{sec:cover}
Although the seed theory is free, twist insertions impose boundary conditions relating fields in different copies. Following \cite{Lunin:2000yv,Lunin:2001pw}, we remove them by mapping the correlator to a covering surface, on which the fields are single-valued. On this covering surface, the remaining correlator is a seed theory correlator and hence simple to evaluate.

Consider first a fixed assignment of the copy labels. If an insertion at $z_i$ contains a cycle of length $n_i$, the covering map $z(t)$ has the following behavior near the corresponding preimage $t_i$,
\begin{align}
    z-z_i = b_i(t-t_i)^{n_i} +O\bigl((t-t_i)^{n_i+1}\bigr)\,, \qquad b_i \neq 0\,.
\end{align}
The $n_i$ copies involved in this cycle are mapped to the ramification point $t_i.$ Copies that are not permuted at $z_i$ are, on the other hand, mapped to different preimages. 
The connected components are determined by the orbits of the group generated by the twist permutations at the insertions. For the three-point coefficients evaluated below, every connected component of the covering surface has genus zero.\footnote{This is the main advantage of using the Gava-Narain method compared to computing the full four-point function and integrating over the insertion of the modulus, as in \eqref{eq:gammadef}. The four-point functions one would compute involve higher genus covering surfaces, on which correlators are significantly more difficult to evaluate than those on genus zero surfaces.} In Table \ref{tab:covers} we list the covering maps we use. Finally, for an untwisted operator, the relevant preimage is determined by the copy on which it acts.

\begin{table}[ht]
\centering
\renewcommand{\arraystretch}{1.3}
\begin{tabular}{c|c|c}
twists of $(O',\phi,\chi)$ & degree & covering map\\
\hline
\hline
$(2,1,2)$ & 2 & $z=t^2$\\
$(2,2,1)$ & 2 & $z=\dfrac{t^2}{2t-1}$\\
$(2,2,3)$ & 3 & $z=(t-1)^2(2t+1)$\\
$(2,3,2)$ & 3 & $z=\dfrac{(t-1)^2(t+2)}{2-3t}$\\
$(2,3,4)$ & 4 & $z=t^2(3t^2-8t+6)$
\end{tabular}
\caption{Covering maps, with $O'$ at $z=0$, $\phi$ at $z=1$ and $\chi$ at $z=\infty$. The untwisted operator sits at the unbranched preimages, $t=\pm1$ in the first row and $t=\infty,\frac12$ in the second.}
\label{tab:covers}
\end{table}

Twisted operators are represented by oscillator states of the seed theory on the cover, acting on the appropriate cover ground state. Since $\Psi^{(I)}(t)=(dz/dt)^{1/2}\Psi^{(I)}(z)$ and $dz/dt\sim n_ib_i(t-t_i)^{n_i-1}$, a fermion acquires the monodromy $(-1)^{n_i-1}$ around $t_i$.
Starting from the NS sector on the base, odd cycles are mapped to NS states on the cover, and even cycles map to Ramond states. Ramond ground states are created by spin fields (which we discuss in detail in Section~\ref{sec:bosonization}). The natural coordinate near an $n_i$-cycle is $(z-z_i)^{1/n_i}=b_i^{1/n_i}(t-t_i)+O\big((t-t_i)^2\big)$. A cover primary of weights $(h^{\rm cov},\widetilde h^{\rm cov})$, therefore, picks up the factor
\begin{align}
    b_i^{-h^{\rm cov}/n_i}\,\bar b_i^{-\widetilde h^{\rm cov}/n_i}\,.
\end{align}
Here and in the cover correlator, fractional powers are taken on the principal branch. When the base insertion or its preimage lies at infinity, the corresponding coordinate is inverted: $z-z_i\to 1/z$ for $z_i=\infty$ and $t-t_i\to 1/t$ for $t_i=\infty$.

When the cover state is a descendant, the subleading terms of the covering map also contribute. With $w=(z-z_i)^{1/n_i}$ and $u=t-t_i$, the modes map to
\begin{align}
    \hspace{-2pt}\partial X_{-m}^{(I)}\longmapsto\oint_{u=0}\frac{du}{2\pi i}\,w(u)^{-m}\,\partial X^{(I)}(u)\,,\quad
    \Psi_{-r}^{(I)}\longmapsto\oint_{u=0}\frac{du}{2\pi i}\,w(u)^{-r-\frac12}\left(\frac{dw}{du}\right)^{\frac12}\Psi^{(I)}(u)\,.
\end{align}
Consider a weight-one current $J$ acting on a cover primary $|q\rangle$ with $J_0|q\rangle=q|q\rangle$. If $b_i^{-1/n_i}w(u)=u+a_iu^2+O(u^3)$, then
\begin{align}\label{eq:current-cover}
    J_{-1}^{(w)}|q\rangle\longmapsto b_i^{-(h_q+1)/n_i}\left(J^{(u)}_{-1}-a_i\,q\right)|q\rangle\,.
\end{align}
For example, $a_i=-\frac13$ at $t=0$ for the $(2,2,3)$ cover map in Table \ref{tab:covers}. Antiholomorphic modes transform in the same way.

The metric induced from the base is not flat in the coordinate $t$, and removing the conformal factor $|dz/dt|^2$ produces a Liouville factor~\cite{Lunin:2000yv}. For the genus-zero covers considered here, it is fixed by the local behavior of the covering map near the ramification points and infinity. After normalizing the twist operators, the regulator-dependent terms cancel, leaving a finite bare-twist factor. This factor is $1$ for the degree-two maps, $3^{3/2}2^{-8/3}$ for the degree-three maps, and $2^{31/12}3^{-15/8}$ for the degree-four map. We include it together with the local-coordinate factors and the correlator evaluated on the cover.

\subsection{Bosonization}\label{sec:bosonization}
It will be convenient to bosonize the fermions in order to evaluate the correlators on the cover. We introduce two free bosons $H^{(1)},H^{(2)}$ for the
left-moving fermions, with
\begin{align}
    H^{(a)}(t)H^{(b)}(t')
    \sim
    -\delta^{ab}\log(t-t')\,,
\end{align}
and write
\begin{align}
    \Psi^{(1)}(t)
    &=
    \eta_1 e^{iH^{(1)}(t)},
    &
    \Psi^{(3)}(t)
    &=
    \eta_1 e^{-iH^{(1)}(t)},
    \nonumber\\
    \Psi^{(2)}(t)
    &=
    \eta_2 e^{iH^{(2)}(t)},
    &
    \Psi^{(4)}(t)
    &=
    \eta_2 e^{-iH^{(2)}(t)},
\end{align}
with
\begin{align}
    :\!\Psi^{(1)}\Psi^{(3)}\!: \,=
    i\partial H^{(1)},
    \qquad
    :\!\Psi^{(2)}\Psi^{(4)}\!: \,=
    i\partial H^{(2)}.
\end{align}
The right-moving fermions are bosonized similarly using $\widetilde H^{(1)},\widetilde H^{(2)}$ and Klein factors
$\eta_3,\eta_4$, with $\{\eta_a,\eta_b\}=2\delta_{ab}$.

The Ramond ground states are represented by spin fields
\begin{align}
    \sigma_R^{k_1k_2;\widetilde k_1\widetilde k_2}
    &=
    \eta_1^{\,k_1+\frac12}
    \eta_2^{\,k_2+\frac12}
    \eta_3^{\,\widetilde k_1+\frac12}
    \eta_4^{\,\widetilde k_2+\frac12}\,
    e^{i\left(
         k_1 H^{(1)}
        +k_2 H^{(2)}
        +\widetilde k_1\widetilde H^{(1)}
        +\widetilde k_2\widetilde H^{(2)}\right)}\,,
\end{align}
where $k_i,\widetilde{k}_i =  \pm \frac{1}{2}$. These satisfy $h^{\rm cov}=\widetilde h^{\rm cov}=\frac14$. Using
\begin{align}\label{eq:twist-weight}
    h=\frac{h^{\rm cov}}{n}+\frac{c}{24}\left(n-\frac1n\right)\,,
\end{align}
the corresponding twist-2 states, therefore, have $h=\widetilde h=\frac12$. The half-modulus on the cover is represented by
\begin{align}\label{eq:Oprime-cover}
    O'(0)
    \longmapsto
    \frac{1}{\sqrt{2}}\left(\partial X^{(3)}\,\sigma_R^{\frac12,-\frac12;-\frac12,-\frac12}
    +
    \partial X^{(4)}\,\sigma_R^{-\frac12,\frac12;-\frac12,-\frac12}\right)(t_0)\,,
\end{align}
where $t_0$ is the ramification point over $z=0$. Similarly,
\begin{align}\label{eq:Oprimedag-cover}
    O'^\dagger(0)
    \longmapsto
    \frac{1}{\sqrt{2}}\left(\partial X^{(1)}\,\sigma_R^{-\frac12,\frac12;\frac12,\frac12}
    -
    \partial X^{(2)}\,\sigma_R^{\frac12,-\frac12;\frac12,\frac12}\right)(t_0)\,.
\end{align}

We denote by $|0\rangle_{\rm R}$ the spin field with the charges specified in each case, and by $|0\rangle_{\rm NS}$ the NS vacuum. Excited states on the cover are written as modes acting on these ground states. A fermionic mode shifts the charge of the vertex operator it acts on, 
\begin{align}
    \Psi^{(1)}(u)\,e^{iqH^{(1)}}(0)
    =
    \eta_1\,u^{q}\,e^{i(q+1)H^{(1)}}(0)\Big(1+u\,i\partial H^{(1)}(0)+O(u^2)\Big)\,.
\end{align}
The modes $\Psi^{(1)}_{-r}$ with $r<q+\frac12$, therefore, annihilate $e^{iqH^{(1)}}$, while $\Psi^{(1)}_{-q-\frac12}$ simply raises the charge to $q+1$. The modes of $\Psi^{(3)}$ instead lower the charge to $q-1$, and $\Psi^{(2)},\Psi^{(4)}$ and the right-moving fermions act in the same way.
After the Klein factors are evaluated, we are left with a correlator of free bosons
\begin{align}\label{eq:bosonized-correlator}
    \Big\langle \prod_{a} e^{i(q_a\cdot H+\widetilde q_a\cdot\widetilde H)}(t_a,\bar t_a)\Big\rangle
    =
    \prod_{a<b}(t_a-t_b)^{q_a\cdot q_b}\,(\bar t_a-\bar t_b)^{\widetilde q_a\cdot\widetilde q_b}\,,
\end{align}
where $H=(H^{(1)},H^{(2)})$ and $\widetilde H=(\widetilde H^{(1)},\widetilde H^{(2)})$, and the operators are ordered as in the correlator. It vanishes unless $\sum_a q_a=\sum_a\widetilde q_a=0$.

\subsection{Selection rules and orbit factors}\label{sec:combinatorics}
We evaluate each three-point coefficient in \eqref{eq:master_GN} for fixed-copy representatives $\check\chi,\check\phi,\check O'$, and then sum over their $S_N$ orbits. For fixed representatives, the correlator vanishes unless the permutation of $\check\chi$ is the product of those of $\check\phi$ and $\check O'$. Since $\check O'$ carries a single 2-cycle, this fixes the possible twists of $\chi$. In particular, it forces $\chi$ and $\phi$ to lie in sectors of opposite parity, i.e., one permutation is even and the other odd. For example, the untwisted and twist-3 sectors are even, and the twist-2 and twist-$(2,1)$ sectors are odd.

If $\phi$ is twisted, this selection rule also allows sectors in which the 2-cycle of $\check O'$ is disjoint from the twisted cycle of $\check\phi$. If this is the case, the cover is disconnected, and the correlator factorizes into contributions from each connected component of the cover. One factor contains $\check O'$, the 2-cycle of $\check\chi$, and whatever $\check\phi$ places on these two copies. This factor always vanishes due to one of the following reasons:
\begin{itemize}
    \item[i)] $\check\phi$ is trivial on the two copies. Then the factor is a two-point function. It is non-zero only if the 2-cycle of $\check\chi$ carries the weights $(1,\frac12)$ of $O'$. Then $h_\chi=h+1$, in contradiction with \eqref{eq:Hpm}.
    \item[ii)] $\check\phi$ excites at least one of the two copies. Recall that $h=1$. The twisted cycle of $\check\phi$ already carries $h\geq\frac12$, so the excitation on the two copies that $O'$ is on has left-moving weight at most $\frac12$ and contains no left-moving boson. Similarly, $h_\chi=1$ forces the 2-cycle of $\check\chi$ onto its ground state. The boson $\partial X$ in $O'$, see \eqref{eq:Oprime-cover}, therefore has nothing to contract with.
\end{itemize}
Hence, for twisted $\phi$, the 2-cycle of $\check O'$ must overlap with the twisted cycle of $\check\phi$. Disconnected covers can still contribute, for example for $\chi$ with twist-$(3,1)$. The corresponding sectors are listed in Table~\ref{tab:selection}, and their covering maps are listed in Table~\ref{tab:covers}.

\begin{table}[ht]
\centering
\renewcommand{\arraystretch}{1.4}\setlength{\tabcolsep}{4pt}\small
\begin{tabular}{c||c|c||c|c|c}
$\phi_k\,\backslash\,\phi_\ell$ & untwisted & $(3)$ & $(2)$ & $(2,1)$ & $(2,1^2)$\\
\hline\hline
untwisted & $(2,1^n)$ & $(2),(2,1)$ & \ding{55} & \ding{55} & \ding{55}\\
\hline
$(3)$ & $(2),(2,1)$ & $(2),(2,1),(4)$ & \ding{55} & \ding{55} & \ding{55}\\
\hline\hline
$(2)$ & \ding{55} & \ding{55} & $(1^n),(3)$ & -- & --\\
\hline
$(2,1)$ & \ding{55} & \ding{55} & -- & $(1^n),(3),(3,1)$ & $(3),(3,1)$\\
\hline
$(2,1^2)$ & \ding{55} & \ding{55} & -- & $(3),(3,1)$ & $(3),(3,1)$
\end{tabular}
\caption{Sectors of the intermediate states $\chi$ that contribute to the lifting matrix between two primaries $\phi_k,\phi_\ell$. Here, $(2,1^n)$ and $(1^n)$ denote a 2-cycle and an untwisted state with $n$ excited copies with $n\le3$. A cross marks pairs excluded by the selection rule, and a dash marks pairs for which no allowed sector contributes.}
\label{tab:selection}
\end{table}

Since the deformation is invariant under $SU(2)_1$ (see Section~\ref{sec:seed}), $\gamma$ is block diagonal in $SU(2)_1$ representations and degenerate within each multiplet. We compute the full matrices, with every component of a multiplet treated separately, so this degeneracy is a non-trivial consistency check of the results in Section~\ref{sec:results}.

With the orbit normalization \eqref{eq:orbit_norm}, each of the three operators is a sum over $S_N$. Since the correlator is invariant under a simultaneous permutation of all three operators, the sum over this diagonal $S_N$ gives a factor of $N!$ and allows us to fix the copies of $\check O'$, leaving
\begin{align}\label{eq:orbit-sum}
    C_{\chi^\dagger\phi\,O'}
    =
    \frac{1}{\sqrt{N!\,A_\chi A_\phi A_{O'}}}
    \sum_{g,g'\in S_N}
    \big\langle
    (g\check\chi)^\dagger(\infty)\,
    (g'\check\phi)(1)\,
    \check O'(0)
    \big\rangle\,,
\end{align}
and similarly for $O'^\dagger$.

Only a finite number of relative placements contribute. In the case that $\check\phi$ contains fermions on several copies, $g\check\phi$ includes the sign from reordering them. If each of the three operators is a single cycle with no excitations on further copies, all non-vanishing terms are equal, and the sum reduces to
\begin{align}
    C_{\chi^\dagger\phi\,O'}
    =
    \frac{\mathcal N_{\chi\phi O'}}{\sqrt{|\mathcal C_\chi|\,|\mathcal C_\phi|\,|\mathcal C_{O'}|}}\,
    \check C_{\chi^\dagger\phi\,O'}\,,
\end{align}
where $|\mathcal C|$ is the number of distinct copy assignments of each field, $\mathcal N_{\chi\phi O'}$ is the number of copy assignments of all three operators with a non-vanishing correlator, and $\check C$ is the coefficient for fixed copies. For example, for the $(2,2,3)$ cover, $|\mathcal C_2|=\frac12N(N-1)$, $|\mathcal C_3|=\frac13N(N-1)(N-2)$ and $\mathcal N_{223}=N(N-1)(N-2)$, so that
\begin{align}
    \frac{\mathcal N^2}{|\mathcal C_2|^2\,|\mathcal C_3|}=\frac{12(N-2)}{N(N-1)}\,.
\end{align}
The entire $N$-dependence of $\gamma$ and of the Gram matrices enters through these factors. Since they are exact in $N$, so are the lifts obtained by diagonalizing $\gamma$. 

\section{Results}\label{sec:results}
We now collect the lifting matrices for $\widetilde j\in\{0,\frac12,1\}$. All results are exact in $N$. In each case, we order the primaries by their twist sector. First we discuss the results for $N\ge4$ for the different values of $\tilde{j}$ in turn. Then in Section \ref{sec:bkz}, we compare our results to those of BKZ \cite{Benjamin:2021zkn}. Finally, we discuss the special cases $N\in\{2,3\}$ in Section \ref{sec:smallN}. The primaries and the intermediate states that contribute are listed in Appendix~\ref{app:sectors}.

\subsection[\texorpdfstring{$\widetilde j=0$}{j=0}]{\texorpdfstring{$\boldsymbol{\widetilde j=0}$}{j=0}}\label{sec:results-zero}
At $\tilde{j}=0,$ the three flavor current primaries couple only to twist-$2$ intermediate states in the $O'^\dag$ channel. We obtain
\begin{align}\label{eq:gamma-zero}
    \delta h=\frac{\pi^2\lambda^2}{4(N-1)}\,\1_3\,.
\end{align}
The three primaries form a triplet of $SU(2)_1,$ and are all lifted by the same amount. This computation is worked out in detail in Appendix~\ref{sec:example}.

\subsection[\texorpdfstring{$\widetilde j=\frac{1}{2}$}{j=1/2}]{\texorpdfstring{$\boldsymbol{\widetilde j=\frac{1}{2}}$}{j=1/2}}\label{sec:results-half}
At $\tilde{j} = \frac{1}{2},$ there are six untwisted, six twist-$2$, and two twist-$(2,1)$ primaries. The untwisted and twisted primaries lie in even and odd sectors respectively. Since $\chi$ and $\phi$ lie in sectors of opposite parity, the twisted and untwisted primaries do not mix, as explained in Section~\ref{sec:combinatorics}. The twist-2 and twist-$(2,1)$ primaries do not mix either, since they form two triplets and a doublet of $SU(2)_1$. After combining all steps of the computation, we get
\begin{align}\label{eq:gamma-half}
    \delta h=\frac{\pi^2\lambda^2}{4(N-1)}
    \begin{pNiceArray}{c|c|c}[first-row,first-col]
     & \text{untwisted} & \text{twist-2} & \text{twist-}(2,1) \\
    \text{untwisted} & \frac12\,\1_6 & 0 & 0 \\
    \hline
    \text{twist-2} & 0 & \1_6 & 0 \\
    \hline
    \text{twist-}(2,1) & 0 & 0 & \frac23\,\1_2
    \end{pNiceArray}.
\end{align}
In units of $\frac{\pi^2 \lambda^2}{4(N-1)}$, the three sectors are lifted as follows:
\begin{itemize}
    \item[] \textbf{Untwisted.} The six primaries form the multiplets $\mathbf 4\oplus\mathbf 2$ and are lifted by $\frac12$. They couple only to twist-2 and twist-$(2,1)$ intermediate states, all in the $O'^\dagger$ channel.
    \item[] \textbf{Twist-2.} The two triplets $\mathbf 3\oplus\mathbf 3$ are lifted by $1$, but through different channels. One triplet couples only to $O'$, and only to untwisted intermediate states. The other couples only to $O'^\dagger$, and receives the following contributions
    \begin{align}\label{eq:tw2-split}
        \frac{\delta h}{\pi^2\lambda^2/(4(N-1))}=\underbrace{\tfrac{N+6}{4N}}_{\text{untwisted}}+\underbrace{\tfrac{3(N-2)}{4N}}_{\text{twist-3}}=1\,.
    \end{align}
    The two contributions add up to exactly the lift~\eqref{eq:gamma-zero} of the flavor currents. This allows the six twist-2 primaries to combine with the three flavor currents into the multiplets $3\,\widetilde\chi_{0,0}$ of Table~\ref{tab:recomb-results} (together with 3 primaries at $\tilde{j}=1$).
    \item[] \textbf{Twist-$\boldsymbol{(2,1)}$.} The doublet $\mathbf 2$ is lifted by $\frac23$, entirely by twist-3 intermediate states. These have $\widetilde h_\chi\geq\frac23$, so they only appear in the $O'^\dagger$ channel, which has $\widetilde h_\chi=1$.
\end{itemize}
The contributions of the individual sectors of $\chi$ are listed in Table~\ref{tab:sectors-half} in Appendix \ref{app:sectors}.

There are also four $\frac{1}{2}$-BPS primaries with $(j,\tilde{j})=(1,\frac{1}{2}),$ two untwisted and two twist-$(2,1).$ These do not lift, as they are protected by supersymmetry. From the perspective of \eqref{eq:master_GN}, the vanishing is due to the fact that all possible OPE coefficients contributing to the anomalous dimensions are zero.

\subsection[\texorpdfstring{$\widetilde j=1$}{j=1}]{\texorpdfstring{$\boldsymbol{\widetilde j=1}$}{j=1}}\label{sec:results-one}

At $\widetilde j=1$, there are nine untwisted, seven twist-3, sixteen twist-$(2,1)$ and four twist-$(2,1^2)$ primaries. They are not orthonormal, so the lifts $\delta h$ are the generalized eigenvalues, given by
\begin{equation}
\det\big(\gamma-\delta h\,G_\phi\big)=0\,,
\label{eq:lifts-def}
\end{equation}
where $G_\phi$ is the Gram matrix of the primaries. The permutations of the untwisted and twist-3 sectors are even, and those of the twist-$(2,1)$ and twist-$(2,1^2)$ sectors are odd. The two pairs, therefore, do not mix with each other. Within each pair, however, there is mixing. We use the multiplet names of Appendix~\ref{app:sectors-one}. The untwisted triplets are~$\mathbf 3_1$,~$\mathbf 3_2$ and~$\mathbf 3_3$, the twist-3 triplet is~$\mathbf 3_{\text{tw-}3}$, and a subscript~$\pm$ marks a multiplet that is lifted only by intermediate states in~$\CH_\pm$. In units of $\frac{\pi^2\lambda^2}{4(N-1)}$, the sectors lift as:
\begin{itemize}
    \item[] \textbf{Untwisted and twist-3.} The sixteen even primaries have
    \begin{align}\label{eq:lifts16}
        \mathbf 3_3:&\qquad 0\,,
        \nonumber\\
        \mathbf 3_1\oplus\mathbf 3_2\oplus\mathbf 3_{\text{tw-}3}:&\qquad
        \underbrace{\tfrac1N,\ \tfrac12}_{\text{untwisted}},\ \underbrace{\tfrac{5N-6}{6N}}_{\text{twist-3}}
        \;\xrightarrow{\ \text{mixing}\ }\; 0,\ \tfrac13,\ 1\,,
        \nonumber\\
        \mathbf 2_+\oplus\mathbf 2_-:&\qquad \tfrac23\,.
    \end{align}
    The untwisted triplet~$\mathbf 3_3$ couples to none of the intermediate states. The triplets~$\mathbf 3_1$ and~$\mathbf 3_2$ mix with the twist-3 triplet through twist-2 intermediate states, and the arrow indicates the change in the lifts due to this mixing. The twist-3 doublet~$\mathbf 2_+$ is lifted through~$O'^\dagger$, and~$\mathbf 2_-$ through~$O'$.
    \item[] \textbf{Twist-$\boldsymbol{(2,1)}$ and twist-$\boldsymbol{(2,1^2)}$.} The twenty odd primaries have
    \begin{align}\label{eq:lifts20}
        (\mathbf 4\oplus\mathbf 2)_+\oplus(\mathbf 4\oplus\mathbf 2)_-:&\qquad \tfrac12\,,
        \nonumber\\
        (\mathbf 3\oplus\mathbf 1)_{(2,1)}\oplus(\mathbf 3\oplus\mathbf 1)_{(2,1^2)}:&\qquad
        \underbrace{\tfrac{4(N-3)}{9(N+6)}}_{\text{twist-}(2,1)},\ \underbrace{\tfrac{4(N-1)^2}{9(N-2)(N+1)}}_{\text{twist-}(2,1^2)}
        \;\xrightarrow{\ \text{mixing}\ }\; 0,\ \tfrac49\,.
    \end{align}
    The twelve primaries in the first line are lifted,~$(\mathbf 4\oplus\mathbf 2)_+$ through~$O'^\dagger$ and~$(\mathbf 4\oplus\mathbf 2)_-$ through~$O'$. The others mix through twist-3 intermediate states, and the arrow again indicates the effect of the mixing.
\end{itemize}
The full matrices $\gamma$ and $G_\phi$ are given in Appendix~\ref{app:sectors-one}. The lifted primaries combine into the long multiplets according to~\eqref{recomb}. The resulting long multiplets, together with their anomalous dimensions, are presented in Table \ref{tab:recomb-results}.

\begin{table}[ht]
\centering
\renewcommand{\arraystretch}{1.3}
\begin{tabular}{@{}c@{\qquad}r@{{}={}}l@{}}
\toprule
$\displaystyle\frac{\delta h}{\pi^2\lambda^2/[4(N-1)]}$
  & \multicolumn{2}{l}{Multiplet recombination \eqref{recomb}} \\
\midrule
$1$
  & $3\,\widetilde\chi_{0,0}$
  & $3\,\widetilde\chi_0+6\,\widetilde\chi_{\frac12}
     +3\,\widetilde\chi_1$ \\
$\frac12$
  & $6\,\widetilde\chi_{\frac12,\frac12}$
  & $6\,\widetilde\chi_{\frac12}+12\,\widetilde\chi_1
     +\textcolor{gray}{6\,\widetilde\chi_{\frac32}}$ \\
$\frac23$
  & $2\,\widetilde\chi_{\frac12,\frac12}$
  & $2\,\widetilde\chi_{\frac12}+4\,\widetilde\chi_1
     +\textcolor{gray}{2\,\widetilde\chi_{\frac32}}$ \\
$\frac13$
  & $3\,\widetilde\chi_{1,1}$
  & $3\,\widetilde\chi_1
     +\textcolor{gray}{6\,\widetilde\chi_{\frac32}
     +3\,\widetilde\chi_2}$ \\
$\frac49$
  & $4\,\widetilde\chi_{1,1}$
  & $4\,\widetilde\chi_1
     +\textcolor{gray}{8\,\widetilde\chi_{\frac32}
     +4\,\widetilde\chi_2}$ \\
\bottomrule
\end{tabular}
\caption{Multiplet recombination at the indicated anomalous dimensions. The terms in grey are predictions for $\widetilde j=\frac32$ and $\widetilde j=2$, the latter for $N\geq5$.}
\label{tab:recomb-results}
\end{table}
Comparison to the total number of~$1/4$-BPS primaries, summarized in Table~\ref{tab:spectrum}, shows that exactly ten primaries remain unlifted at $\widetilde j=1$. This is in agreement with the number of~$1/4$-BPS primaries at the supergravity point~\cite{Benjamin:2016pil}. Furthermore, all 1/4-BPS primaries considered here with $\widetilde j\leq\frac12$ are lifted.

There are also twenty-one~$\frac{1}{2}$-BPS primaries with~$(j,\widetilde j)=(1,1)$: eleven untwisted, one twist-3, one twist-$(2, 2)$, four twist-$(2,1)$ and four twist-$(2,1^2)$. These do not lift; all possible OPE coefficients vanish.

\subsection{Comparison to BKZ}\label{sec:bkz}

Because our computation is a generalization of the one performed by BKZ in \cite{Benjamin:2021zkn}, it is prudent to compare and contrast our results. BKZ computed the untwisted block of the lifting matrix. In the process of computing the full lifting matrix, we noticed an inaccuracy in the way Hermitian conjugation is implemented in the code used by \cite{Benjamin:2021zkn}. This inaccuracy leads to a non-Hermitian lifting matrix (which was not noticed because only the upper triangle of the untwisted block was computed, which was then completed by hand to a Hermitian block). Correctly implementing the Hermitian conjugation does not change the qualitative conclusions of the paper. Moreover, because the lifting matrices for $\tilde{j}\le\frac{1}{2}$ are diagonal, they are unaffected by the issue with Hermitian conjugation. The $\tilde{j}\le\frac{1}{2}$ entries computed in BKZ agree with ours. To ease the comparison for $\tilde{j}=1$, we repeat the eigenvalues one would obtain from just the untwisted block at $\tilde{j}=1$ (cf. \eqref{eq:lifts16})\footnote{These are not true eigenvalues because of the nontrivial mixing with twisted sectors. If, however, one (wrongly) assumes that mixing does not happen, these would be the eigenvalues of the untwisted block.} 
\begin{align}\label{eq:untwlift}
        \mathbf 3_{1}:&\qquad \frac{\pi^2\lambda^2}{4N(N-1)}\,,
        \nonumber\\
                \mathbf 3_{2}:&\qquad \frac{\pi^2\lambda^2}{8(N-1)}\,,\\
        \mathbf 3_{3}:&\qquad 0\,.
        \nonumber
\end{align}

There is a triplet of operators that seemingly obtains an anomalous dimension that scales like $1/N^2$, shown here on the top line. Operator mixing with operators in the twist-3 sector exactly cancels this and all subleading contributions to the full anomalous dimension. Computing only certain blocks of the lifting matrix can thus lead to qualitatively different results! A consequence is that the interpretation of the $1/N^2$ lifting in terms of a binding energy for multi-particle states of protected operators is not applicable to the anomalous dimensions in the sectors we considered in this paper (this is the suggested explanation for the $1/N^2$ anomalous dimensions in BKZ). Hence, at least in the sectors we consider in this work, there is no binding energy for multiparticle states whose single-particle components do not lift.

\subsection[\texorpdfstring{${N=2, \; N=3}$}{N=2,3}]{\texorpdfstring{$\boldsymbol{N=2, \; N=3}$}{N=2,3}}\label{sec:smallN}

The results above hold for $N\geq4$, since all primaries and intermediate states that contribute require at most four copies. Specifically, at $N=2$, only the cycle types $(1)$, $(1^2)$ and $(2)$ exist, and we repeat the computation keeping only these states. The only primaries are then the three flavor currents at $\widetilde j=0$, and the six twist-2 primaries at $\widetilde j=\frac12$. They couple to twist-2 and untwisted intermediate states respectively, and all nine are lifted by the same amount,
\begin{align}\label{eq:gamma-N2}
    \gamma=\frac{\pi^2\lambda^2}{4}\,\1_9\,.
\end{align}
They form three long multiplets $\widetilde\chi_0+2\widetilde\chi_{\frac12}$, the truncated form of~\eqref{recomb} at $N=2$.

This lift was computed in~\cite{Guo:2019ady}. Their deformation is normalized differently from~\eqref{eq:mod}, with $\lambda^2=2\lambda_{\rm GM}^2$. Once this is taken into account, we get the same lift,
\begin{align}
    \delta\Delta=\delta h+\delta\widetilde h=\pi^2\lambda_{\rm GM}^2\,.
\end{align}

At $N=3$, the cycle types $(1)$, $(1^2)$, $(1^3)$, $(2)$, $(2,1)$, and $(3)$ exist. There are then $3$, $14$ and $29$ primaries at $\widetilde j=0,\frac12,1$ respectively, and none at $\widetilde j=\frac32$. At $\widetilde j\leq\frac12$, the primaries and the twist sectors of $\chi$ that contribute are the same as for $N\geq4$, and the lifts are given by~\eqref{eq:gamma-zero} and~\eqref{eq:gamma-half} at $N=3$. At $\widetilde j=1$, there are six untwisted, seven twist-3 and sixteen twist-$(2,1)$ primaries. The triplets~$\mathbf 3_1$ and~$\mathbf 3_2$ need four copies, and a single untwisted triplet~$\mathbf 3_{\rm untw}$ takes their place. The triplet~$\mathbf 3_3$ is the same as for~$N\geq4$. In units of $\frac{\pi^2\lambda^2}{4(N-1)}\vert_{N=3}=\frac{\pi^2\lambda^2}{8}$, they lift as
\begin{align}\label{eq:lifts-N3}
    \mathbf 3_3:&\qquad 0\,,
    \nonumber\\
    \mathbf 3_{\rm untw}\oplus\mathbf 3_{\text{tw-}3}:&\qquad
    \underbrace{\tfrac12}_{\text{untwisted}},\ \underbrace{\tfrac12}_{\text{twist-3}}
    \;\xrightarrow{\ \text{mixing}\ }\; 0,\ 1\,,
    \nonumber\\
    \underbrace{\mathbf 2_+\oplus\mathbf 2_-}_{\text{twist-3}}:&\qquad \tfrac23\,,
    \nonumber\\
    \underbrace{(\mathbf 4\oplus\mathbf 2)_+\oplus(\mathbf 4\oplus\mathbf 2)_-}_{\text{twist-}(2,1)}:&\qquad \tfrac12\,,
    \nonumber\\
    (\mathbf 3\oplus\mathbf 1)_{\text{twist-}(2,1)}:&\qquad 0\,.
\end{align}
Compared to~\eqref{eq:lifts16} and~\eqref{eq:lifts20}, one untwisted triplet and the twist-$(2,1^2)$ primaries are absent. Thus, the lifts $\frac13$ and $\frac49$ do not appear. The lifted primaries combine into the multiplets shown in Table \ref{tab:recomb-N3}. We see that ten primaries remain unlifted at $\widetilde j=1$, similar to our findings at $N\geq4$. The $\frac12$-BPS primaries, four at $\widetilde j=\frac12$ and fifteen at $\widetilde j=1$, are not lifted.

\begin{table}[ht]
\centering
\renewcommand{\arraystretch}{1.3}
\begin{tabular}{@{}c@{\qquad}r@{{}={}}l@{}}
\toprule
$\displaystyle\frac{8 \delta h}{\pi^2\lambda^2}$
  & \multicolumn{2}{l}{Multiplet recombination \eqref{recomb}} \\
\midrule
$1$
  & $3\,\widetilde\chi_{0,0}$
  & $3\,\widetilde\chi_0+6\,\widetilde\chi_{\frac12}
     +3\,\widetilde\chi_1$ \\
$\frac12$
  & $6\,\widetilde\chi_{\frac12,\frac12}$
  & $6\,\widetilde\chi_{\frac12}+12\,\widetilde\chi_1$ \\
$\frac23$
  & $2\,\widetilde\chi_{\frac12,\frac12}$
  & $2\,\widetilde\chi_{\frac12}+4\,\widetilde\chi_1$ \\
\bottomrule
\end{tabular}
\caption{Multiplet recombination at $N=3$. The long multiplets at $\widetilde j=\frac12$ take the truncated form of~\eqref{recomb}.}
\label{tab:recomb-N3}
\end{table}

The $N=3$ lifts of the flavor currents and of the twisted primaries at $\widetilde j=\frac12$ were also computed in~\cite{Zhang:2026jnf}, up to an overall normalization. We fix this normalization with the flavor currents. The six twist-2 primaries then have the same lift as here. For the twist-$(2,1)$ doublet,~\cite{Zhang:2026jnf} finds a lift larger than ours by a factor $\frac43$, that is, $\frac89$ instead of $\frac23$. Our value comes entirely from twist-3 intermediate states; see Table~\ref{tab:sectors-half} in Appendix \ref{app:sectors}. It is confirmed by the recombination in Table \ref{tab:recomb-N3}: the same lift appears at $\widetilde j=1$ for both twist-3 doublets of~\eqref{eq:lifts-N3}. The doublet~$\mathbf 2_+$ receives it from intermediate states with~$\widetilde h_\chi=\frac32$, and~$\mathbf 2_-$ from intermediate states with~$\widetilde h_\chi=\frac12$, in both cases in the sectors~$(2)$ and~$(2,1)$. These three-point coefficients are independent of the one that lifts the twist-$(2,1)$ doublet, but nevertheless yield the same anomalous dimension, as the recombination rule~\eqref{recomb} requires. In the planar limit,~\cite{Zhang:2026jnf} finds a ratio~$\frac{8}{27}$ between the lifts of the corresponding doublet and of the flavor currents, while ours is~$\frac23$ for all~$N$.

\section{Discussion}\label{sec:disc}
In this paper, we studied the fate of the lightest 1/4-BPS states of~$\mathrm{Sym}^N(T^4)$ under the exactly marginal deformation that moves the theory towards the supergravity point. We worked in conformal perturbation theory, at second order in the deformation parameter~$\lambda$. We computed the complete lifting matrices for the primaries with~$h=1~,j=0$ and~$\widetilde{h}=\widetilde{j}\in \{0,\tfrac{1}{2},1\}$. All results are exact in~$N$ and include all pertinent twisted sectors. This generalizes the untwisted-sector computation of~BKZ \cite{Benjamin:2021zkn}. Our main results are summarized below.
    
    Not all operators that can lift do lift. Every 1/4-BPS primary with~$\widetilde{j} \le \tfrac{1}{2}$ is lifted, for all~$N\ge 2$. At~$\widetilde{j}=1$, exactly~$10$ primaries remain unlifted, for every~$N\ge 3 $. This is precisely the number of~$1/4$-BPS primaries with these quantum numbers at the supergravity point~\cite{Benjamin:2017rnd}. Earlier computations~\cite{Burrington:2012yq, Gaberdiel:2015uca, Hampton:2018ygz, Guo:2019ady,Keller:2019suk,Guo:2020gxm, Apolo:2022fya, Guo:2022ifr, Hughes:2023apl, Hughes:2023fot, Gaberdiel:2026jor, Zhang:2026jnf,Benjamin:2021zkn} had suggested the opposite pattern, i.e.~that at finite~$N$, every state that can recombine does so at first non-trivial order. Moreover, mixing between the different twist sectors is essential to reach this conclusion. If the~$\widetilde{j}=1$ lifting matrix is restricted to its untwisted block, one triplet acquires an anomalous dimension~$\pi^2\lambda^2/(4N(N-1))$. Mixing with the twist-$3$ triplet, mediated by twist-$2$ intermediate states, cancels this contribution exactly. The same happens in the odd sector, where mixing between the twist-$(2,1)$ and twist-$(2,1^2)$ primaries turns the~$N$-dependent lifts in~\eqref{eq:lifts20} into~$0$ and~$\tfrac{4}{9}$. Truncating the lifting matrix to a block can therefore give qualitatively wrong answers. This is true even in a large-$N$ expansion, where the eigenvalues in question are subleading. 

    All anomalous dimensions take a simple form. Once mixing is included, every non-zero eigenvalue can be written as
    \begin{equation}
    \delta h=\alpha\,\frac{\pi^2\lambda^2}{4(N-1)}\,, \qquad
    \alpha\in\Big\{\tfrac13,\tfrac49,\tfrac12,\tfrac23,1\Big\}\,.
  \end{equation}
In particular, all of them start at order~$1/N$, in agreement with conjecture~$1$ of~\cite{Chang:2025rqy}. There are no eigenvalues that start at higher inverse powers of $N$ in the sectors we considered. The interpretation of such terms as binding energies of multi-particle states, proposed by BKZ, therefore does not apply here.

The lifted operators recombine into complete long multiplets. This is shown in Table~\ref{tab:recomb-results} for~$N\ge 4$ and Table~\ref{tab:recomb-N3} for~$N=3$. All members of the long multiplets acquire equal anomalous dimensions, as required, even when their lifting is mediated by unrelated operators. For example, the lift of~$\tfrac{2}{3}$ of the twist-$(2,1)$ doublet at~$\widetilde{j}=\tfrac{1}{2}$ comes entirely from twist-$3$ intermediate states, while the same lift of the twist-$3$ doublet~$\mathbf{2}_+$ at~$\widetilde{j}=1$ comes from intermediate states in the sectors $(2)$, $(2,1)$ and, for $N\geq4$, $(4)$. We see this as a strong consistency check on our computations. The gray entries in~Table \ref{tab:recomb-results} are predictions for the lifts of the~$\widetilde{j}=\tfrac32$ and~$\widetilde{j}=2$ members of these multiplets.

We consider $N\in\{2,3\}$ separately. At~$N=2$, the only primaries are the three flavor currents and the six twist-~$2$ primaries. All nine are lifted by the same amount, reproducing the results of~\cite{Guo:2019ady}, as expected. At~$N=3$, one untwisted triplet and the twist~$(2,1^2)$ primaries are absent, and ten primaries remain unlifted at~$\widetilde{j}=1$, as for~$N\ge4$.

Two comments are in order.
\begin{enumerate}
    \item[i)] Our results hold at~$\CO(\lambda^2)$. A vanishing eigenvalue of the second-order lifting matrix does not by itself guarantee that an operator is BPS at generic points on the conformal manifold. The non-renormalization conjecture of~\cite{Chang:2025rqy} states that it does. For the purpose of this work, we assume this conjecture.\footnote{Note, however, that the analogous statement does not hold in the context of $\mathcal{N}=4$ SYM, where in the theory with gauge group SO(7) evidence was found of lifting that only happens at loop level \cite{chang2025violationsdualityclassicalqcohomology}.} With this in mind, we can conclude that, at least in the sectors considered here, the number of BPS states at the generic points agrees with that at the supergravity point.

    \item[ii)] Assuming the non-renormalization conjecture just mentioned, the ten unlifted operators are monotone in the sense of~\cite{Chang:2024zqi,Chang:2025rqy,Chang:2025wgo}. They are BPS for every~$N\ge3$ at which they exist, while all other operators we considered are lifted for all~$N$. This is consistent with~\cite{Chang:2025wgo}, where the lowest-lying fortuitous cohomology classes were found at~$h=2$. Since~$\widetilde{j}=1$ is fixed while~$N$ varies, the unlifted operators lie away from the charge-concentration locus~$\widetilde{j}=N/2$ for all~$N\ge3$. The existence of such non-charge concentrated~$1/4$-BPS operators was recently proven non-perturbatively in~\cite{Belin:2026vfd}. The operators found here are explicit examples. 
\end{enumerate}
We close with some directions for future work.
\begin{itemize}
    \item[]{\bf{  Higher values of~$\boldsymbol{h}$ and~$\boldsymbol{\widetilde{j}}$.}} This is the most natural extension of our computations. It tests the predictions in Table~\ref{tab:recomb-results}, and it shows whether the agreement between the second-order perturbation theory and supergravity~$1/4$-BPS spectra persists beyond the lowest level. At~$h=2$, it should expose the first fortuitous operators and the value of~$N$ at which they lift. Our results show that the full mixing problem must be solved in such a computation; no twist sector can be dropped.

    \item[]{\bf{OPE coefficients and the chiral algebra.}} It would be interesting to compare the coefficients of the unlifted operators with the bootstrap predictions of~\cite{Belin:2026vfd}, and to understand how these operators organize themselves into the chiral algebra there. 

    \item[]{\bf{Other seed theories.}} For the non-linear sigma model on~$K3$, the elliptic genus is non-vanishing, so the comparison to supergravity is direct. The lifting of various operators in this theory was studied in~\cite{Keller:2019suk,Keller:2023ssv}. A second family is given by (tensor products of)~$\CN=(2,2)$ minimal models, which are rational and completely solved. These theories are classified by the existence of a single-trace exactly marginal operator and by the growth of their elliptic genus~\cite{Belin:2020nmp,Benjamin:2022jin}. In~\cite{Benjamin:2022jin}, theories were found with a single-trace twisted-sector marginal operator whose elliptic genus is nevertheless inconsistent with a supergravity point. Comparing the~$1/4$-BPS spectra of such theories with those that do admit a supergravity point could clarify the conditions holographic CFTs must satisfy.
    
\end{itemize}

\section*{Acknowledgments}
We thank Alex Belin, Nathan Benjamin, Christoph Keller, Kyriakos Papadodimas, and Ida Zadeh for discussions. We thank the Mani L. Bhaumik Institute for Theoretical Physics for support. AG was supported by the U.S. Department of Energy through award DE-SC0009937, as well as the Simons Foundation through the Collaboration on Confinement and QCD Strings under award 994302 and through the Collaboration on Global Categorical Symmetries.

\paragraph{Statement of AI use.} The writing of our code was, in part, assisted by ChatGPT Astra and Claude Fable 5.1, and we used these models to perform various consistency checks. The research questions, methodology, algorithm, analysis, and interpretation of the results are our own.  All work was reviewed, verified, and modified by us, and we take full responsibility for the accuracy and content of the paper.

\appendix 
\section{The \texorpdfstring{$\boldsymbol{\CN=(4,4)}$}{N=4,4} superconformal algebra}\label{22sca}

In this appendix we collect our conventions for the small~$\CN=(4,4)$ superconformal algebra, its free-field realization on~$T^4$, and its extension to the contracted large~$\CN=(4,4)$ algebra. The~$\CN=(4,4)$ algebra can be split into a left- and right-moving algebra, which commute with one another. Our conventions mostly follow those of BKZ~\cite{Benjamin:2021zkn} except for an overall sign in the flavor currents, with which the flavor currents satisfy the commutation relations of a standard~$su(2)$ algebra.

\subsection*{The small~\texorpdfstring{$\boldsymbol{\CN=(4,4)}$}{N=(4,4)} algebra}

The small $\CN=(4,4)$ algebra is generated by the stress tensor~$T$, four supercurrents~$G^{\alpha A}$ and three~$SU(2)$ R-currents~$J^a$, with mode expansions
\begin{equation}\label{eq:appA-modes}
    T(z) = \sum_{n\in\Z}\frac{L_n}{z^{n+2}}~,\qquad G^{\alpha A}(z) = \sum_{r\in\Z+\frac{1}{2}}\frac{G^{\alpha A}_r}{z^{r+\frac{3}{2}}}~,\qquad J^a(z) = \sum_{n\in\Z}\frac{J^a_n}{z^{n+1}}~.
\end{equation}
Here~$\alpha\in\{\pm\}$ is a doublet index of the R-symmetry,~$A\in\{1,2\}$ is a doublet index of the outer automorphism, and~$a\in\{3,\pm\}$. For example, we write~$G^{(+1)}$ for the component~$\alpha=+$,~$A=1$. The Virasoro algebra at central charge~$c$ and the action of~$L_m$ on~$G^{\alpha A}_r$ and~$J^a_n$ are the standard ones, see e.g. \cite{Eguchi:1987wf}.
The R-currents form an~$SU(2)$ current algebra at level~$k=\frac{c}{6}$, under which the supercurrents are doublets,
\begin{align}\label{eq:appA-JJ}
    [J^3_m,J^3_n] &= \frac{c}{12}\,m\,\delta_{m+n,0}~,\qquad [J^3_m,J^\pm_n] = \pm J^\pm_{m+n}~,\qquad [J^+_m,J^-_n] = 2J^3_{m+n}+\frac{c}{6}\,m\,\delta_{m+n,0}~,\nonumber\\
    [J^3_m,G^{\pm A}_r] &= \pm\frac{1}{2}\,G^{\pm A}_{m+r}~,\qquad [J^\pm_m,G^{\mp A}_r] = G^{\pm A}_{m+r}~,\qquad [J^\pm_m,G^{\pm A}_r] = 0~.
\end{align}
Finally, the supercurrents satisfy
\begin{align}\label{eq:appA-GG}
    \{G^{+A}_r,G^{-B}_s\} &= \epsilon^{AB}\Big(L_{r+s}+(r-s)\,J^3_{r+s}+\frac{c}{6}\Big(r^2-\frac{1}{4}\Big)\delta_{r+s,0}\Big)~,\nonumber\\
    \{G^{+A}_r,G^{+B}_s\} &= -\epsilon^{AB}\,(r-s)\,J^+_{r+s}~,\qquad \{G^{-A}_r,G^{-B}_s\} = \epsilon^{AB}\,(r-s)\,J^-_{r+s}~,
\end{align}
with~$\epsilon^{12}=-\epsilon^{21}=1$. Hermitian conjugation acts as
\begin{align}\label{eq:appA-dagger}
    L_n^\dagger &= L_{-n}~,\qquad \big(J^3_n\big)^\dagger = J^3_{-n}~,\qquad \big(J^\pm_n\big)^\dagger = J^\mp_{-n}~,\nonumber\\
    \big(G^{(+1)}_r\big)^\dagger &= G^{(-2)}_{-r}~,\qquad \big(G^{(+2)}_r\big)^\dagger = -\,G^{(-1)}_{-r}~.
\end{align}

A superconformal primary~$\phi$ is annihilated by all modes~$L_n, J^a_n, G^{\alpha A}_r$ with~$n,r>0$. We denote its weight by~$h$, its R-symmetry spin by~$j$, and its~$J^3_0$ eigenvalue by~$j^3$. It follows from~\eqref{eq:appA-GG} and~\eqref{eq:appA-dagger} that, for~$A\in\{1,2\}$, we have
\begin{equation}\label{eq:appA-bound}
    \big\|G^{+A}_{-\frac{1}{2}}\phi\big\|^2 = \big(h-j^3\big)\big\|\phi\big\|^2~,\qquad \big\|G^{-A}_{-\frac{1}{2}}\phi\big\|^2 = \big(h+j^3\big)\big\|\phi\big\|^2\qquad\Longrightarrow\qquad h\geq j~.
\end{equation}
This unitarity bound is saturated by chiral primaries. Their highest-weight component~$j^3=j$ is annihilated by~$G^{+A}_{-\frac{1}{2}}$, and their lowest-weight component~$j^3=-j$ is annihilated by~$G^{-A}_{-\frac{1}{2}}$. These are known as the shortening conditions. Primaries with~$h=j$ hence generate short representations, while primaries with~$h>j$ generate long ones.

\subsection*{Free-field realization on~\texorpdfstring{$\boldsymbol{T^4}$}{T4}}

The nonlinear sigma model on $T^4$ has~$c=6$. Its field content can be summarized by the left-moving fields ~$\partial X^{(I)}$ and~$\Psi^{(I)}$, with~$I=1,\dots,4$. At times we use the following assignment for conjugate labels:~$\bar 1=3$,~$\bar 2 = 4$. The modes of the fields are introduced in~\eqref{eq:s3-modes}--\eqref{eq:s3-dagger}, which are equivalent to the OPEs
\begin{equation}\label{eq:appA-OPE}
    \partial X^{(I)}(z)\,\partial X^{(J)}(w)\sim\frac{\delta^{I\bar J}}{(z-w)^2}~,\qquad \Psi^{(I)}(z)\,\Psi^{(J)}(w)\sim\frac{\delta^{I\bar J}}{z-w}~.
\end{equation}
The following combinations of the fundamental fields form the generators of the small~$\CN=(4,4)$ algebra that the theory satisfies,
\begin{equation}\label{eq:appA-G}
\begin{gathered}
T=\partial X^{(1)}\partial X^{(3)}+\partial X^{(2)}\partial X^{(4)}+\frac12\sum_{I=1}^{4}\partial\Psi^{(I)}\Psi^{(\bar I)}\,,\\[4pt]
\begin{alignedat}{2}
G^{(+1)}&=\Psi^{(1)}\partial X^{(3)}+\Psi^{(2)}\partial X^{(4)}\,,&\qquad G^{(-1)}&=\Psi^{(3)}\partial X^{(4)}-\Psi^{(4)}\partial X^{(3)}\,,\\
G^{(+2)}&=-\Psi^{(1)}\partial X^{(2)}+\Psi^{(2)}\partial X^{(1)}\,,&\qquad G^{(-2)}&=\Psi^{(3)}\partial X^{(1)}+\Psi^{(4)}\partial X^{(2)}\,,\\[4pt]
J^{+}&=-\Psi^{(1)}\Psi^{(2)}\,,&\qquad J^{-}&=\Psi^{(3)}\Psi^{(4)}\,,
\end{alignedat}\\
J^{3}=\frac12\Big(\Psi^{(1)}\Psi^{(3)}+\Psi^{(2)}\Psi^{(4)}\Big)\,.
\end{gathered}
\end{equation}
It is not hard to see that indeed these combinations of fields obey~\eqref{eq:appA-JJ}--\eqref{eq:appA-dagger} with~$c=6$. The bosons are neutral under the R-symmetry, and the fermions~$\Psi^{(1)},\Psi^{(2)}$ satisfy~$j^3=+\frac{1}{2}$, while~$\Psi^{(3)},\Psi^{(4)}$ satisfy~$j^3=-\frac{1}{2}$.

The fermion bilinears that are neutral under the R-symmetry generate a second~$SU(2)$ current algebra with level one, which commutes with~$J^a$. The currents of this~$SU(2)$ are known as the flavor currents~$\hat J^a$ of~\eqref{eq:s3-flavor}, with~$\big(\hat J^{\pm}_n\big)^\dagger = \hat J^{\mp}_{-n}$.
The modes of the flavor currents obey the first line of~\eqref{eq:appA-JJ} with~$J^a\mapsto\hat J^a$ and~$c=6$. The fermions~$\Psi^{(1)},\Psi^{(4)}$ have~$\hat J^3_0$ charge~$+\frac{1}{2}$, and~$\Psi^{(2)},\Psi^{(3)}$ have charge~$-\frac{1}{2}$. The flavor currents do not commute with the supercurrents, and they do not map them into each other. For instance,
\begin{equation}\label{eq:appA-flavorG}
    [\hat J^+_m,G^{(+1)}_r] = \big(\Psi^{(1)}\partial X^{(4)}\big)_{m+r}~,
\end{equation}
which is not a supercurrent. The flavor currents are therefore not part of the superconformal algebra.

The symmetries are most transparent in bispinor notation. We define
\begin{align}\label{eq:appA-bispinor}
    \Psi^{+\dot 1} &= \Psi^{(1)}~, & \Psi^{+\dot 2} &= \Psi^{(2)}~, & \Psi^{-\dot 1} &= -\Psi^{(4)}~, & \Psi^{-\dot 2} &= \Psi^{(3)}~,\nonumber\\
    \partial X^{1\dot 1} &= -\partial X^{(4)}~, & \partial X^{1\dot 2} &= \partial X^{(3)}~, & \partial X^{2\dot 1} &= -\partial X^{(1)}~, & \partial X^{2\dot 2} &= -\partial X^{(2)}~.
\end{align}
In this notation, the OPEs~\eqref{eq:appA-OPE} take the covariant form
\begin{equation}\label{eq:appA-cov}
    \Psi^{\alpha\dot A}(z)\,\Psi^{\beta\dot B}(w)\sim\frac{\epsilon^{\alpha\beta}\epsilon^{\dot A\dot B}}{z-w}~,\quad \partial X^{A\dot A}(z)\,\partial X^{B\dot B}(w)\sim\frac{\epsilon^{AB}\epsilon^{\dot A\dot B}}{(z-w)^2}~,
\end{equation}
 and the supercurrents~\eqref{eq:appA-G} can be written as
 \eq{
\quad G^{\alpha A} = \epsilon_{\dot A\dot B}\,\Psi^{\alpha\dot A}\,\partial X^{A\dot B}~,
}
with~$\epsilon^{+-}=\epsilon^{12}=\epsilon^{\dot 1\dot 2}=\epsilon_{\dot 1\dot 2}=1$. The index~$\alpha$ is rotated by the R-symmetry, and~$\dot A$ is rotated by~$SU(2)_1$, which acts on bosons and fermions and leaves~$G^{\alpha A}$ invariant. The index~$A$ is rotated by~$SU(2)_2$, which acts on the bosons alone and forms the outer automorphism of the algebra. The flavor currents~\eqref{eq:s3-flavor} rotate the index~$\dot A$ of only the fermions,~$[\hat J^+_m,\Psi^{\alpha\dot 2}_r]=\Psi^{\alpha\dot 1}_{m+r}$.

\subsection*{The contracted large~\texorpdfstring{$\boldsymbol{\CN=(4,4)}$}{N=(4,4)} algebra}

The contracted large~$\CN=(4,4)$ algebra is obtained by adjoining the free fields~$\partial X^{A\dot A}$ and~$\Psi^{\alpha\dot A}$ to the small~$\CN=(4,4)$ algebra. They are conformal, resp. superconformal primaries of weight~$1$ and~$\frac{1}{2}$. The bosons are R-symmetry singlets, and the fermions are R-symmetry doublets,~$[J^\mp_m,\Psi^{\pm\dot A}_r]=\Psi^{\mp\dot A}_{m+r}$. The supercurrents exchange the bosons and fermions according to
\begin{equation}\label{eq:appA-large}
    \{G^{\alpha A}_r,\Psi^{\beta\dot B}_s\} = \epsilon^{\alpha\beta}\,\partial X^{A\dot B}_{r+s}~,\qquad [G^{\alpha A}_r,\partial X^{B\dot B}_m] = \epsilon^{AB}\,m\,\Psi^{\alpha\dot B}_{r+m}~.
\end{equation}
For example,~$\{G^{(+1)}_r,\Psi^{(3)}_s\}=\partial X^{(3)}_{r+s}$ and~$[G^{(+1)}_r,\partial X^{(1)}_m]=-m\,\Psi^{(1)}_{r+m}$.

In~$\text{Sym}^N(T^4)$ all generators are diagonal over the~$N$ copies. The relations above continue to hold with~$c=6N$, and with an additional factor of~$N$ on the right-hand sides of~\eqref{eq:s3-algebra} and~\eqref{eq:appA-OPE}.

We end by commenting on the Hermitian conjugate~$\phi^\dagger$ of an operator~$\phi$, when it appears in correlation functions. It is defined so that~$\langle 0|\phi^\dagger(\infty)$ is the bra conjugate to the state~$\phi(0)|0\rangle$, with~$\phi^\dagger(\infty)=\lim_{z\to\infty}z^{2h}\bar z^{2\widetilde h}\phi^\dagger(z,\bar z)$. On states built from modes, conjugation acts as in~\eqref{eq:s3-dagger} and reverses the order of all modes, including fermion modes that act on different copies. For the twist fields we use~$(\sigma^{--})^\dagger=\sigma^{++}$, so that~$O'=G^{(+1)}_{-\frac{1}{2}}\sigma^{--}$ becomes ~$O'^\dagger = -G^{(-2)}_{-\frac{1}{2}}\sigma^{++}$ under Hermitian conjugation.

\section{Example: lifting of flavor currents}\label{sec:example}

At $\widetilde j=0$, the three primaries are the flavor-current primaries~\eqref{eq:s3-flavorprimary}, with fixed-copy representatives
\begin{align}\label{eq:flavor-reps}
    \check\phi^3 &= \frac{1}{\sqrt{2}} \Big( -\Psi^{(1),1}_{-\frac12}\Psi^{(3),1}_{-\frac12}
    +\Psi^{(2),1}_{-\frac12}\Psi^{(4),1}_{-\frac12}
    + \Psi^{(1),1}_{-\frac12}\Psi^{(3),2}_{-\frac12}
    -\Psi^{(2),1}_{-\frac12}\Psi^{(4),2}_{-\frac12} \Big) |0\rangle\,,  \nonumber \\
    \check\phi^- &= \Big( \Psi^{(2),1}_{-\frac12}\Psi^{(3),1}_{-\frac12}
    - \Psi^{(2),1}_{-\frac12}\Psi^{(3),2}_{-\frac12} \Big) |0\rangle\,,  \nonumber \\
    \check\phi^+ &= \Big( \Psi^{(1),1}_{-\frac12}\Psi^{(4),1}_{-\frac12}
    - \Psi^{(1),1}_{-\frac12}\Psi^{(4),2}_{-\frac12} \Big) |0\rangle\,.
\end{align}
In each line, the terms with both fermions on copy 1 act on a single copy, and the others act on copies 1 and 2. The orbit normalization~\eqref{eq:orbit_norm} gives $A_\phi=(N-1)!+(N-2)!=N(N-2)!$. Let us compute the coefficients of $\phi^-$.

Since the primaries have $(h,\widetilde h)=(1,0)$, the channel $\CH_-$ would require $\widetilde h_\chi=-\frac12$ and is empty. Thus only the $O'^\dagger$ channel contributes, and the intermediate states have
\begin{align}
    (h_\chi,\widetilde h_\chi;j^3_\chi,\widetilde j^{\,3}_\chi)=\big(1,\tfrac12;0,\tfrac12\big)\,.
\end{align}
By Table~\ref{tab:selection}, they lie in the twist-$2$ and twist-$(2,1)$ sectors. The sector $(2,1)$ contains $4$ states, and all their coefficients vanish. The twist-$2$ sector contains $12$ states. Only the $4$ states $\big(\partial X^{(I)}_{-1}\Psi^{(J)}_{0}\widetilde\Psi^{(1)}_{0}\widetilde\Psi^{(2)}_{0}|0\rangle_{\rm R}\big)_{\rm cov}$ with $I,J\in\{1,2\}$ couple to the primaries, and $\phi^-$ couples to one of them. On the cover of its 2-cycle, it is
\begin{align}\label{eq:flavor-chi}
    |\check\chi\rangle_{\rm cov}=\partial X_{-1}^{(2)}\Psi_0^{(2)}\,\widetilde\Psi_0^{(1)}\widetilde\Psi_0^{(2)}|0\rangle_{\rm R}\,,
\end{align}
where $|0\rangle_{\rm R}$ has the charges $(-\frac12,-\frac12;-\frac12,-\frac12)$. The zero modes raise them to $(-\frac12,\frac12;\frac12,\frac12)$. The cover weights are $(h^{\rm cov},\widetilde h^{\rm cov})=(\frac54,\frac14)$, and~\eqref{eq:twist-weight} gives $(h_\chi,\widetilde h_\chi)=(1,\frac12)$. At infinity, the conjugate state is represented by
\begin{align}
    \check\chi^\dagger(\infty)
    \longmapsto
    \lim_{t\to\infty}
    t^{\frac52}\bar t^{\frac12}\,
    \partial X^{(4)}(t)\,
    \sigma_R^{\frac12,-\frac12;-\frac12,-\frac12}(t,\bar t)\,.
\end{align}

Let us fix the 2-cycle of $\check O'^\dagger$ to be $(12)$ and use the $(2,1,2)$ map $z=t^2$ of Table~\ref{tab:covers}. The half-modulus then sits at $t=0$ and $\check\chi^\dagger$ at $t=\infty$. The primary sits at $z=1$, with the preimage $t=1$ on copy 1 and $t=-1$ on copy 2. For this map, $b_i=1$ at both ramification points and the bare-twist factor is $1$. Thus only the fermions in the primary give local factors. Since $dz/dt=2t$, they are
\begin{align}
    \Psi^{(2),1}(1)\Psi^{(3),1}(1)
    &\longmapsto
    \frac12\,\Psi^{(2)}(1)\Psi^{(3)}(1)\,,
    \nonumber\\
    \Psi^{(2),1}(1)\Psi^{(3),2}(1)
    &\longmapsto
    2^{-\frac12}(-2)^{-\frac12}\,\Psi^{(2)}(1)\Psi^{(3)}(-1)\,.
\end{align}
Here $(-2)^{-\frac12}=2^{-\frac12}(-1)^{-\frac12}$ is taken on the branch of $t^{-\frac12}$ at $t=-1$, since $(dz/dt)^{-\frac12}\Psi(t)$ is single-valued in $z$ in the presence of the spin field at $t=0$. The fermions of $\phi^-$ carry the charges $(-1,1)$ under $(H^{(1)},H^{(2)})$. Charge conservation thus selects the second term of~\eqref{eq:Oprimedag-cover}, whose spin field has the charges $(\frac12,-\frac12;\frac12,\frac12)$. The one-copy piece is then
\begin{align}
    \check C^{\text{1-copy}}
    = -\frac{1}{2\sqrt{2}}
    \lim_{t\to\infty}
    t^{\frac52}\bar t^{\frac12}
    \Big\langle
    &\partial X^{(4)}(t)\,
    \sigma_R^{\frac12,-\frac12;-\frac12,-\frac12}(t,\bar t)
    \nonumber\\
    &\times
    \Psi^{(2)}(1)\Psi^{(3)}(1)\,
    \partial X^{(2)}(0)\,
    \sigma_R^{\frac12,-\frac12;\frac12,\frac12}(0)
    \Big\rangle\,,
\end{align}
where the prefactor is the $-\frac{1}{\sqrt2}$ of~\eqref{eq:Oprimedag-cover} times the local factor $\frac12$. At $(t,1,0)$, the charges are $(\frac12,-1,\frac12)$ under $H^{(1)}$ and $(-\frac12,1,-\frac12)$ under $H^{(2)}$. Including the Klein factors,~\eqref{eq:bosonized-correlator} gives
\begin{align}
    \Big\langle
    \sigma_R^{\frac12,-\frac12;-\frac12,-\frac12}(t,\bar t)\,
    \Psi^{(2)}(1)\Psi^{(3)}(1)\,
    \sigma_R^{\frac12,-\frac12;\frac12,\frac12}(0)
    \Big\rangle
    =
    \frac{t^{\frac12}}{t-1}\,\bar t^{-\frac12}\,,
\end{align}
while the bosons contract to $\langle\partial X^{(4)}(t)\,\partial X^{(2)}(0)\rangle=t^{-2}$. Thus
\begin{align}
    \check C^{\text{1-copy}}
    =
    -\frac{1}{2\sqrt2}
    \lim_{t\to\infty}
    \frac{t^{\frac52}\bar t^{\frac12}}{t^2}\,
    \frac{t^{\frac12}}{t-1}\,\bar t^{-\frac12}
    =-\frac{1}{2\sqrt2}\,.
\end{align}
In the two-copy piece, $\Psi^{(3)}$ sits at $t=-1$ instead. This changes the $H^{(1)}$ correlator to $(t+1)^{-\frac12}t^{\frac14}(-1)^{-\frac12}$. With the relative minus sign of this term in $\check\phi^-$, we find
\begin{align}
    \check C^{\text{2-copy}}
    &= \frac{1}{\sqrt{2}}\,2^{-\frac12}(-2)^{-\frac12}
    \lim_{t\to\infty}
    \frac{t^{\frac52}\bar t^{\frac12}}{t^2}\,
    (-1)^{-\frac12}\,\frac{t^{\frac12}}{\sqrt{(t-1)(t+1)}}\,\bar t^{-\frac12}
    =-\frac{1}{2\sqrt2}\,,
\end{align}
where $(-1)^{-\frac12}(-1)^{-\frac12}=-1$, since both factors are on the same branch. The two pieces are thus equal.

In the sum over $g$ in~\eqref{eq:orbit-sum}, only the $2(N-2)!$ elements that map the 2-cycle of $\check\chi$ to $(12)$ contribute. In the sum over $g'$, the one-copy piece contributes whenever $g'$ maps copy 1 into $\{1,2\}$. There are $2(N-1)!$ such elements. The two-copy piece requires in addition that $g'$ maps copy 2 into $\{1,2\}$, which leaves $2(N-2)!$ elements. With $A_\chi=A_{O'^\dagger}=2(N-2)!$, we obtain
\begin{align}
    C_{\chi^\dagger\phi^-O'^\dagger}
    &=
    \frac{2(N-2)!}{\sqrt{N!\,A_\chi A_\phi A_{O'^\dagger}}}
    \Big(2(N-1)!\,\check C^{\text{1-copy}}+2(N-2)!\,\check C^{\text{2-copy}}\Big)
    \nonumber\\
    &=
    -\Big(\frac{\sqrt{N-1}}{\sqrt2\,N}+\frac{1}{\sqrt2\,N\sqrt{N-1}}\Big)
    =-\frac{1}{\sqrt{2(N-1)}}\,.
\end{align}
The other two primaries lift similarly. They couple to the states
\begin{align}
    \phi^+&:\quad\partial X^{(1)}_{-1}\Psi^{(1)}_0\widetilde\Psi_0^{(1)}\widetilde\Psi_0^{(2)}|0\rangle_{\rm R}\,,
    \nonumber\\
    \phi^3&:\quad\partial X^{(2)}_{-1}\Psi^{(1)}_0\widetilde\Psi_0^{(1)}\widetilde\Psi_0^{(2)}|0\rangle_{\rm R}\,,
    \quad\partial X^{(1)}_{-1}\Psi^{(2)}_0\widetilde\Psi_0^{(1)}\widetilde\Psi_0^{(2)}|0\rangle_{\rm R}\,,
\end{align}
where the two states of $\phi^3$ contribute equally. Since these states are orthonormal and each couples to a single primary,~\eqref{eq:master_GN} gives
\begin{align}
    \gamma=\frac{\pi^2\lambda^2}{4(N-1)}\,\1_3\,.
\end{align}

\section{Example: lifting with a twist-\texorpdfstring{$\boldsymbol{3}$}{3} intermediate state}\label{sec:example3}

A twist-3 intermediate state requires a cover of degree three, with a non-trivial bare-twist factor and non-trivial local coordinates at all three insertions. Let us take the first twist-2 primary at $\widetilde j=\frac12$ in~\eqref{eq:jbar_12_tw2_prims},
\begin{align}
    |\check\phi^1\rangle_{\rm cov}
    =
    \partial X_{-1}^{(4)}\Psi_0^{(1)}\,\widetilde\Psi_0^{(1)}\widetilde\Psi_0^{(2)}|0\rangle_{\rm R}\,,
\end{align}
with $|0\rangle_{\rm R}$ as in~\eqref{eq:flavor-chi}. The cover charges are $(\frac12,-\frac12;\frac12,\frac12)$, the cover weights are $(\frac54,\frac14)$, and $A_\phi=2(N-2)!$. In the $O'^\dagger$ channel, the intermediate states have
\begin{align}
    (h_\chi,\widetilde h_\chi;j^3_\chi,\widetilde j^{\,3}_\chi)=(1,1;0,1)\,.
\end{align}
By~\eqref{eq:twist-weight}, a twist-3 state on the NS cover has $h=\frac{h^{\rm cov}}{3}+\frac23$. Thus both cover weights equal $1$. The right-moving part is then fixed to $\widetilde\Psi^{(1)}_{-\frac12}\widetilde\Psi^{(2)}_{-\frac12}|0\rangle_{\rm NS}$, and the left-moving part is $\partial X^{(I)}_{-1}$ or a neutral fermion bilinear. This gives the eight states
\begin{align}\label{eq:tw3-chis}
    |\check\chi\rangle_{\rm cov}
    &=
    \partial X^{(I)}_{-1}\,
    \widetilde\Psi^{(1)}_{-\frac12}\widetilde\Psi^{(2)}_{-\frac12}|0\rangle_{\rm NS}\,,
    &&
    I=1,\ldots,4\,,
    \nonumber\\
    |\check\chi\rangle_{\rm cov}
    &=
    \Psi^{(I)}_{-\frac12}\Psi^{(J)}_{-\frac12}\,
    \widetilde\Psi^{(1)}_{-\frac12}\widetilde\Psi^{(2)}_{-\frac12}|0\rangle_{\rm NS}\,,
    &&
    I=1,2\,,\ J=3,4\,,
\end{align}
with $A_\chi=3(N-3)!$.

Only one of the states~\eqref{eq:tw3-chis} couples to $\phi^1$. The two terms of~\eqref{eq:Oprimedag-cover} carry the left-moving charges $(-\frac12,\frac12)$ and $(\frac12,-\frac12)$. With the first term, $\phi^1$ and $O'^\dagger$ have total left-moving charge zero, and $\chi$ must be neutral. A bosonic $\chi$ then gives a three-point function of the free bosons $\partial X^{(4)}$, $\partial X^{(1)}$ and $\partial X^{(I)}$, which vanishes. A neutral fermion bilinear leaves $\langle\partial X^{(4)}\partial X^{(1)}\rangle$, which also vanishes. With the second term, the total charge is $(1,-1)$. Only $\Psi^{(1)}\Psi^{(4)}$ can absorb it, and the bosons contract as $\langle\partial X^{(4)}\partial X^{(2)}\rangle\neq0$. The only contributing state is thus
\begin{align}
    |\check\chi\rangle_{\rm cov}
    =
    \Psi^{(1)}_{-\frac12}\Psi^{(4)}_{-\frac12}\,
    \widetilde\Psi^{(1)}_{-\frac12}\widetilde\Psi^{(2)}_{-\frac12}|0\rangle_{\rm NS}\,,
\end{align}
whose conjugate at infinity is represented by
\begin{align}
    \check\chi^\dagger(\infty)
    \longmapsto
    \lim_{t\to\infty}
    t^{2}\bar t^{2}\,
    e^{i\left(-H^{(1)}+H^{(2)}-\widetilde H^{(1)}-\widetilde H^{(2)}\right)}(t,\bar t)\,.
\end{align}

The $(2,2,3)$ map of Table~\ref{tab:covers} is
\begin{align}
    z=(t-1)^2(2t+1)\,,
    \qquad
    z-1=t^2(2t-3)\,,
    \qquad
    \frac{dz}{dt}=6t(t-1)\,.
\end{align}
The half-modulus sits at the double point $t=1$ over $z=0$, the primary at the double point $t=0$ over $z=1$, and $\check\chi^\dagger$ at the triple point $t=\infty$. The simple preimages $t=-\frac12$ and $t=\frac32$ carry the copy outside the respective 2-cycle, on which $O'^\dagger$ and $\phi^1$ act trivially. Near the three ramification points, the map is
\begin{align}
    z&=3u^2\Big(1+\tfrac23u\Big)\,,
    &
    u&=t-1\,,
    &
    b_{O}&=3\,,
    &
    a_{O}&=\tfrac13\,,
    \nonumber\\
    z-1&=-3t^2\Big(1-\tfrac23t\Big)\,,
    &
    &
    &
    b_{\phi}&=-3\,,
    &
    a_{\phi}&=-\tfrac13\,,
    \nonumber\\
    \frac1z&=\frac{s^3}{2}\Big(1+\tfrac32s+O(s^2)\Big)\,,
    &
    s&=\frac1t\,,
    &
    b_{\chi}&=\tfrac12\,.
\end{align}
Since $b_O=3$ and $b_\phi=-3$, the bare-twist factor is
\begin{align}
    Z_{(2,2,3)}
    =
    \Big(2^{-\frac{c}{4}}\,3^{-\frac{c}{24}}\Big)^2
    \Big(3^{\frac{c}{3}}\,2^{\frac{c}{18}}\Big)
    =
    3^{\frac32}\,2^{-\frac83}\,,
\end{align}
at $c=6$. The last factor comes from $z\simeq2t^3$ at infinity.

The subleading coefficients $a_O$ and $a_\phi$ do not contribute. Since the bosonic modes act on states without momentum, $q=0$ in~\eqref{eq:current-cover}. For the fermion zero modes, we have
\begin{align}
    w^{-\frac12}\Big(\frac{dw}{du}\Big)^{\frac12}=u^{-\frac12}\Big(1+\tfrac{a_i}{2}\,u+O(u^2)\Big)\,,
\end{align}
and the correction term gives the positive mode $\Psi_1$, which annihilates the spin field. The same holds for the zero mode in~\eqref{eq:Oprimedag-cover}. Thus the local factors are just $b^{-h^{\rm cov}/n}\,\bar b^{-\widetilde h^{\rm cov}/n}$. Up to phases, they are
\begin{align}
    3^{-\frac{1}{2}\left(\frac54+\frac14\right)}=3^{-\frac34}
    \quad\text{at }t=0\text{ and }t=1\,,
    \qquad
    \left(\frac12\right)^{-\frac13(1+1)}=2^{\frac23}
    \quad\text{at }t=\infty\,.
\end{align}
Together with the bare-twist factor, they give
\begin{align}
    3^{\frac32}\,2^{-\frac83}\cdot3^{-\frac34}\cdot3^{-\frac34}\cdot2^{\frac23}=\frac14\,.
\end{align}

The second term of~\eqref{eq:Oprimedag-cover} sits at $t=1$. The charges under $(H^{(1)},H^{(2)})$ are then $(-1,1)$ at $t$, $(\frac12,-\frac12)$ at $0$ and $(\frac12,-\frac12)$ at $1$. Under both $\widetilde H^{(1)}$ and $\widetilde H^{(2)}$, they are $-1$, $\frac12$ and $\frac12$. By~\eqref{eq:bosonized-correlator},
\begin{align}
    \Big\langle e^{i(-H^{(1)}+H^{(2)})}(t)\,e^{\frac i2(H^{(1)}-H^{(2)})}(0)\,e^{\frac i2(H^{(1)}-H^{(2)})}(1)\Big\rangle
    =
    \frac{(-1)^{\frac12}}{t\,(t-1)}\,,
\end{align}
and the same holds for the right-movers. The bosons contract to $\langle\partial X^{(4)}(0)\,\partial X^{(2)}(1)\rangle=1$. Up to a phase, we thus find
\begin{align}
    \check C_{\chi^\dagger\phi^1O'^\dagger}
    =
    \frac{1}{\sqrt2}\cdot\frac14\cdot
    \lim_{t\to\infty}t^2\bar t^2\,\frac{1}{t(t-1)}\,\frac{1}{\bar t(\bar t-1)}
    =
    \frac{1}{4\sqrt2}\,.
\end{align}

The three operators carry a 3-cycle and two 2-cycles. By~\eqref{eq:orbit-sum} and the counting of Section~\ref{sec:combinatorics}, every non-vanishing placement gives the same value. Up to a phase, this gives
\begin{align}
    C_{\chi^\dagger\phi^1O'^\dagger}
    =
    \sqrt{\frac{12(N-2)}{N(N-1)}}\cdot\frac{1}{4\sqrt2}
    =
    \sqrt{\frac{3(N-2)}{8N(N-1)}}\,.
\end{align}
Since this is the only twist-3 state that couples to $\phi^1$,~\eqref{eq:master_GN} gives the twist-3 contribution to its lift,
\begin{align}
    \gamma^{(3)}_{11}=\frac{3\pi^2\lambda^2(N-2)}{16N(N-1)}\,.
\end{align}
The untwisted sectors $(1)$ and $(1,1)$ add $\frac{\pi^2\lambda^2}{16N}+\frac{7\pi^2\lambda^2}{16N(N-1)}$, and the total is $\gamma_{11}=\frac{\pi^2\lambda^2}{4(N-1)}$.

The other twisted primaries at $\widetilde j=\frac12$ work in the same way. The primary $\phi^2$ couples only to the state with $\Psi^{(2)}\Psi^{(3)}$, and its twist-3 contribution equals that of $\phi^1$. The primary $\phi^3$ couples to the two neutral states with $\Psi^{(1)}\Psi^{(3)}$ and $\Psi^{(2)}\Psi^{(4)}$, which contribute equally. Its twist-3 contribution is again $\frac{3\pi^2\lambda^2(N-2)}{16N(N-1)}$. The primaries $\phi^4,\phi^5,\phi^6$ do not couple to any twist-3 state. Each of the two twist-$(2,1)$ primaries in~\eqref{eq:jbar_12_tw21_prims} couples to a single bosonic state of~\eqref{eq:tw3-chis}, namely $\partial X^{(1)}_{-1}$ for $\phi^1$ and $\partial X^{(2)}_{-1}$ for $\phi^2$. Up to a phase, the coefficient is $\frac{1}{\sqrt{3(N-1)}}$. Here the cover terms and the terms with a fermion on the untwisted copy both contribute, and the coefficient is their sum. Altogether, six of the eight states in~\eqref{eq:tw3-chis} couple.

\section{Primaries and intermediate states}\label{app:sectors}
We list here the primaries at $\widetilde j=\frac12$ and $\widetilde j=1$, and the twist sectors of the intermediate states that contribute to their lifts. As in Section~\ref{sec:results}, contributions to the lifts are given in units of $\frac{\pi^2\lambda^2}{4(N-1)}$.

\subsection*{\texorpdfstring{$\boldsymbol{\widetilde j=\frac12}$}{j=1/2}}\label{app:sectors-half}

The intermediate states have $(h_\chi,\widetilde h_\chi;j^3_\chi,\widetilde j^{\,3}_\chi)=(1,1;0,1)$ in $\CH_+$ and $(1,0;0,0)$ in $\CH_-$. Since $\widetilde h_\chi=0$ in $\CH_-$, only untwisted states with a right-moving vacuum enter this channel.

The six untwisted primaries dress the flavor-current primaries~\eqref{eq:flavor-reps} with a right-moving fermion,
\begin{align}\label{eq:jbar_12_untw_prims}
    \check\phi^{a,I}
    =
    \Big(
    \widetilde\Psi_{-\frac12}^{(I),1}
    +\widetilde\Psi_{-\frac12}^{(I),2}
    -2\,\widetilde\Psi_{-\frac12}^{(I),3}
    \Big)\check\phi^a\,,
    \qquad
    a\in\{3,+,-\}\,,
    \qquad
    I=1,2\,,
\end{align}
with $A_\phi=N^2(N-3)!$. The six twist-2 primaries are built on the Ramond cover of a 2-cycle. With $|\widetilde\sigma\rangle\equiv\widetilde\Psi_0^{(1)}\widetilde\Psi_0^{(2)}|0\rangle_{\rm R}$, they are
\begin{align}\label{eq:jbar_12_tw2_prims}
    |\check\phi^1\rangle_{\rm cov}
    &=
    \partial X_{-1}^{(4)}\Psi_0^{(1)}|\widetilde\sigma\rangle\,,
    &
    |\check\phi^4\rangle_{\rm cov}
    &=
    \partial X_{-1}^{(2)}\Psi_0^{(2)}|\widetilde\sigma\rangle\,,
    \nonumber\\
    |\check\phi^2\rangle_{\rm cov}
    &=
    \partial X_{-1}^{(3)}\Psi_0^{(2)}|\widetilde\sigma\rangle\,,
    &
    |\check\phi^5\rangle_{\rm cov}
    &=
    \tfrac{1}{\sqrt2}\big(\partial X_{-1}^{(1)}\Psi_0^{(2)}+\partial X_{-1}^{(2)}\Psi_0^{(1)}\big)|\widetilde\sigma\rangle\,,
    \nonumber\\
    |\check\phi^3\rangle_{\rm cov}
    &=
    \tfrac{1}{\sqrt2}\big(\partial X_{-1}^{(3)}\Psi_0^{(1)}-\partial X_{-1}^{(4)}\Psi_0^{(2)}\big)|\widetilde\sigma\rangle\,,
    &
    |\check\phi^6\rangle_{\rm cov}
    &=
    \partial X_{-1}^{(1)}\Psi_0^{(1)}|\widetilde\sigma\rangle\,,
\end{align}
with $A_\phi=2(N-2)!$. They form the two triplets $\phi^{1,2,3}$ and $\phi^{4,5,6}$ of Section~\ref{sec:results-half}. The two twist-$(2,1)$ primaries combine a 2-cycle with a left-moving fermion, which sits either on the cover or on an untwisted copy. With $|\widetilde\sigma\rangle$ on the copies $(12)$ and a fermion on copy 3, they are
\begin{align}\label{eq:jbar_12_tw21_prims}
    \check\phi^1_{(2,1)}
    &=
    -\Psi^{(1)}_{-1}|\widetilde\sigma\rangle
    +\Psi^{(4)}_{-1}\Psi^{(1)}_0\Psi^{(2)}_0|\widetilde\sigma\rangle
    +\sqrt2\,|\widetilde\sigma\rangle\otimes\Psi^{(1),3}_{-\frac12}|0\rangle
    -\sqrt2\,\Psi^{(1)}_0\Psi^{(2)}_0|\widetilde\sigma\rangle\otimes\Psi^{(4),3}_{-\frac12}|0\rangle\,,
    \nonumber\\
    \check\phi^2_{(2,1)}
    &=
    \Psi^{(2)}_{-1}|\widetilde\sigma\rangle
    +\Psi^{(3)}_{-1}\Psi^{(1)}_0\Psi^{(2)}_0|\widetilde\sigma\rangle
    -\sqrt2\,|\widetilde\sigma\rangle\otimes\Psi^{(2),3}_{-\frac12}|0\rangle
    -\sqrt2\,\Psi^{(1)}_0\Psi^{(2)}_0|\widetilde\sigma\rangle\otimes\Psi^{(3),3}_{-\frac12}|0\rangle\,,
\end{align}
where each term stands for its unnormalized orbit sum. They form the doublet $\mathbf 2$.

The contributions of the twist sectors of $\chi$ are listed in Table~\ref{tab:sectors-half}.

\begin{table}[ht]
\centering
\renewcommand{\arraystretch}{1.5}
\begin{tabular}{cccc|cccc}
twist & channel & states & contributing & $\phi^{a,I}$ & $\phi^{1,2,3}$ & $\phi^{4,5,6}$ & $\phi^{1,2}_{(2,1)}$\\
\hline
$(1)$ & $\CH_+$ & $8$ & $4$ & & $\frac{N-1}{4N}$ & & \\
$(1,1)$ & $\CH_+$ & $64$ & $28$ & & $\frac{7}{4N}$ & & \\
$(2)$ & $\CH_+$ & $24$ & $8$ & $\frac{N-2}{2N}$ & & & \\
$(2,1)$ & $\CH_+$ & $40$ & $8$ & $\frac1N$ & & & \\
$(3)$ & $\CH_+$ & $8$ & $6$ & & $\frac{3(N-2)}{4N}$ & & $\frac23$\\
\hline
$(1)$ & $\CH_-$ & $8$ & $4$ & & & $\frac{N-1}{N}$ & \\
$(1,1)$ & $\CH_-$ & $4$ & $4$ & & & $\frac1N$ & \\
\hline
\multicolumn{4}{c|}{total} & $\frac12$ & $1$ & $1$ & $\frac23$
\end{tabular}
\caption{Lifting matrix at $\widetilde j=\frac12$, split by the twist sector of $\chi$, in units of $\frac{\pi^2\lambda^2}{4(N-1)}$. For each twist sector and channel, we give the number of states of $\chi$, how many of them contribute, and the part of~\eqref{eq:master_GN} that comes from these states on each primary. All parts are diagonal. Since the primaries are orthonormal, the totals are the lifts $\delta h$ of~\eqref{eq:gamma-half}. The sector $(3)$ is worked out in Appendix~\ref{sec:example3}.}
\label{tab:sectors-half}
\end{table}

The coefficients of the two untwisted $\frac12$-BPS primaries vanish for all $20$ states of the sectors $(2)$, $(2,1)$ and $(2,1^2)$ in $\CH_+$. The two twist-$(2,1)$ $\frac12$-BPS primaries are
\begin{align}\label{eq:jbar_12_tw21_halfBPS}
    \check\phi^{1}_{\frac12\text{-BPS}}
    &=
    -\tfrac{1}{\sqrt2}\,\Psi^{(1)}_{-1}\Psi^{(1)}_0\Psi^{(2)}_0|\widetilde\sigma\rangle
    +\Psi^{(1)}_0\Psi^{(2)}_0|\widetilde\sigma\rangle\otimes\Psi^{(1),3}_{-\frac12}|0\rangle\,,
    \nonumber\\
    \check\phi^{2}_{\frac12\text{-BPS}}
    &=
    -\tfrac{1}{\sqrt2}\,\Psi^{(2)}_{-1}\Psi^{(1)}_0\Psi^{(2)}_0|\widetilde\sigma\rangle
    +\Psi^{(1)}_0\Psi^{(2)}_0|\widetilde\sigma\rangle\otimes\Psi^{(2),3}_{-\frac12}|0\rangle\,,
\end{align}
in the notation of~\eqref{eq:jbar_12_tw21_prims}. Their coefficients vanish for all states of the sectors $(1^n)$ and $(3)$.

\subsection*{\texorpdfstring{$\boldsymbol{\widetilde j=1}$}{j=1}}\label{app:sectors-one}

The intermediate states have $(h_\chi,\widetilde h_\chi;j^3_\chi,\widetilde j^{\,3}_\chi)=(1,\frac32;0,\frac32)$ in $\CH_+$ and $(1,\frac12;0,\frac12)$ in $\CH_-$. Each $SU(2)_1$ multiplet of primaries is labeled by its representation, with a subscript $\pm$ if it is lifted only in $\CH_\pm$.

\subsubsection*{Untwisted and twist-3 primaries}
The nine untwisted primaries form three triplets $\mathbf 3_1$, $\mathbf 3_2$ and $\mathbf 3_3$. Their components of lowest $SU(2)_1$ weight are
\begin{align}\label{eq:jbar_1_untw_prims}
    \check\phi^-_{\mathbf 3_1}
    &=
    \big(\Psi^{(2),1}_{-\frac12}-\Psi^{(2),2}_{-\frac12}\big)
    \big(\Psi^{(3),1}_{-\frac12}-\Psi^{(3),2}_{-\frac12}\big)
    \big(\widetilde\Psi^{(1),1}_{-\frac12}-\widetilde\Psi^{(1),3}_{-\frac12}\big)
    \big(\widetilde\Psi^{(2),1}_{-\frac12}+\widetilde\Psi^{(2),3}_{-\frac12}-2\,\widetilde\Psi^{(2),4}_{-\frac12}\big)|0\rangle\,,
    \nonumber\\
    \check\phi^-_{\mathbf 3_2}
    &=
    \big(\Psi^{(2),1}_{-\frac12}-\Psi^{(2),2}_{-\frac12}\big)
    \big(\Psi^{(3),1}_{-\frac12}-\Psi^{(3),2}_{-\frac12}\big)
    \big(\widetilde\Psi^{(1),1}_{-\frac12}-\widetilde\Psi^{(1),3}_{-\frac12}\big)
    \big(\widetilde\Psi^{(2),4}_{-\frac12}-\widetilde\Psi^{(2),2}_{-\frac12}\big)|0\rangle\,,
    \nonumber\\
    \check\phi^-_{\mathbf 3_3}
    &=
    \big(\Psi^{(1),1}_{-\frac12}\Psi^{(3),2}_{-\frac12}+\Psi^{(2),1}_{-\frac12}\Psi^{(4),2}_{-\frac12}\big)
    \big(\widetilde\Psi^{(2),1}_{-\frac12}-\widetilde\Psi^{(2),3}_{-\frac12}\big)
    \big(\widetilde\Psi^{(2),3}_{-\frac12}-\widetilde\Psi^{(2),2}_{-\frac12}\big)|0\rangle\,,
\end{align}
with $A_{\mathbf 3_1}=P\,(N-4)!$, $A_{\mathbf 3_2}=2(N-1)(N-2)(N-4)!$ and $A_{\mathbf 3_3}=2N(N-3)!$, where $P=N^3-N^2-2N+4$. The other components follow by acting with $SU(2)_1$. Equal components of $\mathbf 3_1$ and $\mathbf 3_2$ overlap, $\sum_{g\in S_N}\langle g\check\phi_{\mathbf 3_1}|\check\phi_{\mathbf 3_2}\rangle=-4(N-1)(N-4)!$. The seven twist-3 primaries are built on the NS cover of a 3-cycle,
\begin{align}\label{eq:jbar_1_tw3_prims}
    |\check\phi_{\mathbf 3_{\text{tw-}3}}\rangle_{\rm cov}
    &\in\Big\{
    \Psi_{-\frac12}^{(1)}\Psi_{-\frac12}^{(4)},\ 
    \tfrac{1}{\sqrt2}\big(\Psi_{-\frac12}^{(2)}\Psi_{-\frac12}^{(4)}-\Psi_{-\frac12}^{(1)}\Psi_{-\frac12}^{(3)}\big),\ 
    \Psi_{-\frac12}^{(2)}\Psi_{-\frac12}^{(3)}
    \Big\}\,
    \widetilde\Psi_{-\frac12}^{(1)}\widetilde\Psi_{-\frac12}^{(2)}|0\rangle_{\rm NS}\,,
    \nonumber\\
    |\check\phi_{\mathbf 2_+}\rangle_{\rm cov}
    &\in\Big\{
    \partial X^{(4)}_{-1},\ -\partial X^{(3)}_{-1}
    \Big\}\,
    \widetilde\Psi_{-\frac12}^{(1)}\widetilde\Psi_{-\frac12}^{(2)}|0\rangle_{\rm NS}\,,
    \nonumber\\
    |\check\phi_{\mathbf 2_-}\rangle_{\rm cov}
    &\in\Big\{
    \partial X^{(1)}_{-1},\ \partial X^{(2)}_{-1}
    \Big\}\,
    \widetilde\Psi_{-\frac12}^{(1)}\widetilde\Psi_{-\frac12}^{(2)}|0\rangle_{\rm NS}\,,
\end{align}
with $A_\phi=3(N-3)!$. They form the triplet $\mathbf 3_{\text{tw-}3}$ and the two doublets $\mathbf 2_+$ and $\mathbf 2_-$. The contributions of the twist sectors of $\chi$ are listed in Table~\ref{tab:sectors-one-16}. The twist-4 states use the $(2,3,4)$ map of Table~\ref{tab:covers}.
\begin{table}[ht]
\centering
\renewcommand{\arraystretch}{1.5}\setlength{\tabcolsep}{4pt}\small
\begin{tabular}{cccc|ccccc}
twist & channel & states & contributing & $\mathbf 3_1$ & $\mathbf 3_2$ & $\mathbf 3_{\text{tw-}3}$ & $\mathbf 2_+$ & $\mathbf 2_-$\\
\hline
$(2)$ & $\CH_+$ & $12$ & $8$ & $\frac{(N-3)(N-2)^3}{4NP}$ & $\frac{N-3}{2N(N-1)}$ & $\frac{N-2}{12N}$ & $\frac{2(N-2)}{27N}$ & \\
$(2,1)$ & $\CH_+$ & $84$ & $40$ & $\frac{2(N-3)(N^2-2N+2)}{NP}$ & $\frac{2(N-3)}{N(N-1)(N-2)}$ & $\frac{2}{3N}$ & $\frac{52}{27N}$ & \\
$(2,1^2)$ & $\CH_+$ & $48$ & $4$ & $\frac{4(N-1)^2}{NP}$ & $\frac{2}{N(N-1)(N-2)}$ & & & \\
$(4)$ & $\CH_+$ & $2$ & $2$ & & & & $\frac{16(N-3)}{27N}$ & \\
\hline
$(2)$ & $\CH_-$ & $12$ & $8$ & $\frac{(N-3)(N-2)(N+2)^2}{4NP}$ & $\frac{N-3}{2N(N-1)}$ & $\frac{3(N-2)}{4N}$ & & $\frac{2(N-2)}{3N}$\\
$(2,1)$ & $\CH_-$ & $4$ & $4$ & & & & & $\frac{4}{3N}$\\
\hline
\multicolumn{4}{c|}{total} & $\frac{(N-2)(N^3+N^2-4)}{2NP}$ & $\frac1N$ & $\frac{5N-6}{6N}$ & $\frac23$ & $\frac23$
\end{tabular}
\caption{Diagonal entries of the lifting matrix of the untwisted and twist-3 primaries at $\widetilde j=1$, split by the twist sector of $\chi$, in units of $\frac{\pi^2\lambda^2}{4(N-1)}$. The columns are as in Table~\ref{tab:sectors-half}. None of the states contributes to $\mathbf 3_3$. The totals give the diagonal of~\eqref{eq:gamma16}. The off-diagonal entries come from the sectors $(2)$ and $(2,1)$ in $\CH_+$ and $(2)$ in $\CH_-$, and between $\mathbf 3_1$ and $\mathbf 3_2$ also from $(2,1^2)$.}
\label{tab:sectors-one-16}
\end{table} 
The triplet $\mathbf 3_3$ has vanishing entries in $\gamma$ and is orthogonal to the other primaries. In the basis $(\mathbf 3_1,\mathbf 3_2,\mathbf 3_{\text{tw-}3},\mathbf 2_+,\mathbf 2_-)$ of the other thirteen, the lifting matrix and the Gram matrix are given by
\begin{align}\label{eq:gamma16}
    \setlength{\arraycolsep}{2.4pt}
    \hspace{-40pt}\gamma=\frac{\pi^2\lambda^2}{4N(N-1)}
    \begin{pNiceArray}{cc|ccc}[first-row,first-col,last-col]
    \CodeBefore
        \rectanglecolor{red!9}{1-3}{2-3}
        \rectanglecolor{red!9}{3-1}{3-2}
    \Body
     & \Block{1-2}<\small>{\text{untwisted}} & & \Block{1-3}<\small>{\text{twist-3}} & & & \\
    \Block{2-1}<\small>{\text{untwisted}} & \frac{(N-2)(N^3+N^2-4)}{2P} & -\frac{2\sqrt2\,\sqrt{N-1}}{\sqrt{(N-2)P}} & -\frac{\sqrt{N-3}\,(N^2+2N-4)}{2\sqrt{3P}} & 0 & 0 & {\scriptstyle \mathbf 3_1}\\
     & -\frac{2\sqrt2\,\sqrt{N-1}}{\sqrt{(N-2)P}} & 1 & -\sqrt{\frac{2(N-1)(N-3)}{3(N-2)}} & 0 & 0 & {\scriptstyle \mathbf 3_2}\\
    \hline
    \Block{3-1}<\small>{\text{twist-3}} & -\frac{\sqrt{N-3}\,(N^2+2N-4)}{2\sqrt{3P}} & -\sqrt{\frac{2(N-1)(N-3)}{3(N-2)}} & \frac{5N-6}{6} & 0 & 0 & {\scriptstyle \mathbf 3_{\text{tw-}3}}\\
     & 0 & 0 & 0 & \frac{2N}{3} & 0 & {\scriptstyle \mathbf 2_+}\\
     & 0 & 0 & 0 & 0 & \frac{2N}{3} & {\scriptstyle \mathbf 2_-}
    \end{pNiceArray},
\end{align}
and
\begin{align}\label{eq:G16}
    G_\phi=
    \begin{pNiceArray}{cc|ccc}[first-row,first-col,last-col,margin]
     & \Block{1-2}<\small>{\text{untwisted}} & & \Block{1-3}<\small>{\text{twist-3}} & & & \\
    \Block{2-1}<\small>{\text{untwisted}} & 1 & -\frac{2\sqrt2\,\sqrt{N-1}}{\sqrt{(N-2)P}} & 0 & 0 & 0 & {\scriptstyle \mathbf 3_1}\\
     & -\frac{2\sqrt2\,\sqrt{N-1}}{\sqrt{(N-2)P}} & 1 & 0 & 0 & 0 & {\scriptstyle \mathbf 3_2}\\
    \hline
    \Block{3-1}<\small>{\text{twist-3}} & 0 & 0 & 1 & 0 & 0 & {\scriptstyle \mathbf 3_{\text{tw-}3}}\\
     & 0 & 0 & 0 & 1 & 0 & {\scriptstyle \mathbf 2_+}\\
     & 0 & 0 & 0 & 0 & 1 & {\scriptstyle \mathbf 2_-}
    \end{pNiceArray}.
\end{align}
Each entry multiplies the identity on the multiplets of its row and column. Entries that mix the untwisted and twist-3 sectors are shaded.

\subsubsection*{Twist-\texorpdfstring{$\boldsymbol{(2,1)}$}{(2,1)} and twist-\texorpdfstring{$\boldsymbol{(2,1^2)}$}{(2,1^2)} primaries}
The sixteen twist-$(2,1)$ and four twist-$(2,1^2)$ primaries are obtained from the twisted primaries at $\widetilde j=\frac12$ by adding a right-moving fermion $\widetilde\Psi^{(K)}$, $K=1,2$. With the 2-cycle on the copies $(12)$, this is the replacement
\begin{align}\label{eq:jbar_1_tw21_dressing}
    |\widetilde\sigma\rangle
    \;\longrightarrow\;
    \big(\widetilde\Psi^{(K)}_{-1}-\sqrt2\,\widetilde\Psi^{(K),c}_{-\frac12}\big)|\widetilde\sigma\rangle
\end{align}
in~\eqref{eq:jbar_12_tw2_prims} and~\eqref{eq:jbar_12_tw21_prims}, where the first term acts on the cover and the second on the copy $c$. With $c=3$, it turns $\phi^{1,2,3}$ and $\phi^{4,5,6}$ into the multiplets $(\mathbf 4\oplus\mathbf 2)_+$ and $(\mathbf 4\oplus\mathbf 2)_-$ of twist $(2,1)$. Applied to $\phi^{1,2}_{(2,1)}$, it gives $(\mathbf 3\oplus\mathbf 1)_{(2,1)}$ for $c=3$ and $(\mathbf 3\oplus\mathbf 1)_{(2,1^2)}$ for $c=4$. Equal components of these two multiplets overlap through all terms except those in which both fermions sit outside the cover. This gives the off-diagonal entry of~\eqref{eq:G20}. The contributions of the twist sectors of $\chi$ are listed in Table~\ref{tab:sectors-one-20}.

\begin{table}[ht]
\centering
\renewcommand{\arraystretch}{1.5}\setlength{\tabcolsep}{4pt}\small
\begin{tabular}{cccc|cccc}
twist & channel & states & contributing & $(\mathbf 4\oplus\mathbf 2)_+$ & $(\mathbf 4\oplus\mathbf 2)_-$ & $(\mathbf 3\oplus\mathbf 1)_{(2,1)}$ & $(\mathbf 3\oplus\mathbf 1)_{(2,1^2)}$\\
\hline
$(1^2)$ & $\CH_+$ & $48$ & $24$ & $\frac{3(N-2)}{2N^2}$ & & & \\
$(1^3)$ & $\CH_+$ & $96$ & $48$ & $\frac{3}{N^2}$ & & & \\
$(3)$ & $\CH_+$ & $16$ & $12$ & $\frac{(N-3)^2}{2N^2}$ & & $\frac{4(N-3)^2}{9N(N+6)}$ & $\frac{4(N-3)(N-1)^2}{9N(N-2)(N+1)}$\\
$(3,1)$ & $\CH_+$ & $16$ & $12$ & $\frac{3(N-3)}{2N^2}$ & & $\frac{4(N-3)}{3N(N+6)}$ & $\frac{4(N-1)^2}{3N(N-2)(N+1)}$\\
\hline
$(1)$ & $\CH_-$ & $16$ & $8$ & & $\frac{(N-1)(N-2)}{2N^2}$ & & \\
$(1^2)$ & $\CH_-$ & $32$ & $24$ & & $\frac{3(N-2)}{2N^2}$ & & \\
$(1^3)$ & $\CH_-$ & $8$ & $8$ & & $\frac{2}{N^2}$ & & \\
\hline
\multicolumn{4}{c|}{total} & $\frac12$ & $\frac12$ & $\frac{4(N-3)}{9(N+6)}$ & $\frac{4(N-1)^2}{9(N-2)(N+1)}$
\end{tabular}
\caption{Diagonal entries of the lifting matrix of the twist-$(2,1)$ and twist-$(2,1^2)$ primaries at $\widetilde j=1$, as in Table~\ref{tab:sectors-one-16}. The totals give the diagonal of~\eqref{eq:gamma20}.}
\label{tab:sectors-one-20}
\end{table}

On $(\mathbf 3\oplus\mathbf 1)_{(2,1)}$ and $(\mathbf 3\oplus\mathbf 1)_{(2,1^2)}$, the sectors $(3)$ and $(3,1)$ give the fractions $\frac{N-3}{N}$ and $\frac3N$ of the whole block of~\eqref{eq:gamma20}, including the off-diagonal entry. Both contributions are of rank one. In the basis $\big((\mathbf 4\oplus\mathbf 2)_+,(\mathbf 4\oplus\mathbf 2)_-,(\mathbf 3\oplus\mathbf 1)_{(2,1)},(\mathbf 3\oplus\mathbf 1)_{(2,1^2)}\big)$, the lifting matrix and the Gram matrix are
\begin{align}\label{eq:gamma20}
    \hspace{-10pt}\gamma=\frac{\pi^2\lambda^2}{4(N-1)}
    \begin{pNiceArray}{ccc|c}[first-row,first-col,last-col,margin]
    \CodeBefore
        \rectanglecolor{red!9}{3-4}{3-4}
        \rectanglecolor{red!9}{4-3}{4-3}
    \Body
     & \Block{1-3}<\small>{\text{twist-}(2,1)} & & & \Block{1-1}<\small>{\text{twist-}(2,1^2)} & \\
    \Block{3-1}<\small>{\text{twist-}(2,1)} & \frac12 & 0 & 0 & 0 & {\scriptstyle(\mathbf 4\oplus\mathbf 2)_+}\\
     & 0 & \frac12 & 0 & 0 & {\scriptstyle(\mathbf 4\oplus\mathbf 2)_-}\\
     & 0 & 0 & \frac{4(N-3)}{9(N+6)} & \frac{4(N-1)\sqrt{N-3}}{9\sqrt{(N-2)(N+1)(N+6)}} & {\scriptstyle(\mathbf 3\oplus\mathbf 1)_{(2,1)}}\\
    \hline
    \Block{1-1}<\small>{\text{twist-}(2,1^2)} & 0 & 0 & \frac{4(N-1)\sqrt{N-3}}{9\sqrt{(N-2)(N+1)(N+6)}} & \frac{4(N-1)^2}{9(N-2)(N+1)} & {\scriptstyle(\mathbf 3\oplus\mathbf 1)_{(2,1^2)}}
    \end{pNiceArray},
\end{align}
and
\begin{align}\label{eq:G20}
    G_\phi=
    \begin{pNiceArray}{ccc|c}[first-row,first-col,last-col,margin]
     & \Block{1-3}<\small>{\text{twist-}(2,1)} & & & \Block{1-1}<\small>{\text{twist-}(2,1^2)} & \\
    \Block{3-1}<\small>{\text{twist-}(2,1)} & 1 & 0 & 0 & 0 & {\scriptstyle(\mathbf 4\oplus\mathbf 2)_+}\\
     & 0 & 1 & 0 & 0 & {\scriptstyle(\mathbf 4\oplus\mathbf 2)_-}\\
     & 0 & 0 & 1 & \frac{\sqrt{N-3}\,(N+2)}{\sqrt{(N-2)(N+1)(N+6)}} & {\scriptstyle(\mathbf 3\oplus\mathbf 1)_{(2,1)}}\\
    \hline
    \Block{1-1}<\small>{\text{twist-}(2,1^2)} & 0 & 0 & \frac{\sqrt{N-3}\,(N+2)}{\sqrt{(N-2)(N+1)(N+6)}} & 1 & {\scriptstyle(\mathbf 3\oplus\mathbf 1)_{(2,1^2)}}
    \end{pNiceArray}.
\end{align}

There are also twenty-one $\frac12$-BPS primaries with $(j,\widetilde j)=(1,1)$. Eleven are untwisted, one is twist-3, and one is twist-$(2^2)$. Their coefficients vanish for all $50$ states of the sectors $(2)$, $(2,1)$, $(2,1^2)$ and $(2,1^3)$. The other eight are four twist-$(2,1)$ and four twist-$(2,1^2)$ primaries. Their coefficients vanish for all $84$ states of the sectors $(1^n)$, $(3)$, $(3,1)$, $(2^2)$ and $(2^2,1)$. Thus, none of the $\frac12$-BPS primaries is lifted.

\bibliographystyle{ytphys}
\bibliography{refs}
\end{document}